\documentclass[11pt]{article} 

\usepackage[margin=3cm]{geometry}
\usepackage{amsmath,amssymb,graphicx}

\newcommand{\FBOX}{\hspace*{\fill}$\rule{0.17cm}{0.17cm}$}
\newcommand{\BB}{\hspace*{\fill}$\rule{0.17cm}{0.17cm}\
\rule{0.17cm}{0.17cm}$}

\newcommand{\ol}{\overline} \newcommand{\ul}{\underline}

\newcommand{\elll}{\mbox{\boldmath $l$\unboldmath}}

\def\eref#1{(\ref{#1})}

\newcommand{\Proof}{\noindent {\bf Proof.  }}

\newtheorem{THM}{THEOREM}[section] \newtheorem{LEMMA}[THM]{Lemma}
\newtheorem{CLAIM}[THM]{Claim} \newtheorem{COR}[THM]{Corollary}
\newtheorem{PROP}[THM]{Proposition}

\newtheorem{Thm}[THM]{Theorem} \newcommand{\Theorem}{\begin{Thm}}
\newcommand{\eTh}{\end{Thm}}

\newtheorem{CON}[THM]{Conjecture}
\newcommand{\Conjecture}{\begin{CON}} \newcommand{\eCon}{\end{CON}}

\newtheorem{CONS}[THM]{} \newcommand{\Conjectures}{\begin{CONS}}
\newcommand{\eCons}{\end{CONS}}

\newcommand{\eq}{\begin{equation}} \newcommand{\eeq}{\end{equation}}

\newcommand{\THEOREM}{\begin{THM}} \newcommand{\eT}{\end{THM}}
\newcommand{\Lemma}{\begin{LEMMA}} \newcommand{\eL}{\end{LEMMA}}
\newcommand{\Claim}{\begin{CLAIM}} \newcommand{\eCl}{\end{CLAIM}}
\newcommand{\Corollary}{\begin{COR}} \newcommand{\eCo}{\end{COR}}

\newcommand{\Proposition}{\begin{PROP}} \newcommand{\eP}{\end{PROP}}

\newtheorem{EXAM}{Example}[section]

\newcommand{\Example}{\begin{EXAM}} \newcommand{\eXa}{\end{EXAM}}

\newtheorem{EX}[THM]{Exercise}

\newcommand{\Exercise}{\begin{EX}} \newcommand{\eEx}{\end{EX}}
\newtheorem{EXS}[THM]{} \newcommand{\Exercises}{\begin{EXS}}
\newcommand{\eExs}{\end{EXS}}

\newtheorem{PROBLEM}[THM]{Problem}
\newcommand{\Problem}{\begin{PROBLEM}}
\newcommand{\ePm}{\end{PROBLEM}}

\newtheorem{PROB}[THM]{} 
\newcommand{\Prob}{\begin{PROB}}
\newcommand{\eProb}{\end{PROB}}

\newtheorem{PROBLEMS}[THM]{}
\newcommand{\Problems}{\begin{PROBLEMS}}
\newcommand{\ePms}{\end{PROBLEMS}}

\newtheorem{OPROBLEM}[THM]{Research problem}
\newcommand{\Open}{\begin{OPROBLEM}} 
\newcommand{\eO}{\end{OPROBLEM}}

\newtheorem{REMARK}[THM]{Remark}
\newcommand{\Remark}{\begin{REMARK}} 
\newcommand{\eRe}{\end{REMARK}}

\newtheorem{COMMENT}{Comment}[section]

\newcommand{\Comment}{\begin{COMMENT}} 
\newcommand{\eCom}{\end{COMMENT}}

\newtheorem{TODO}{Todo}[section] \newcommand{\Todo}{\begin{TODO}}
\newcommand{\eTo}{\end{TODO}}

\newtheorem{MEMO}{Memo}[section] \newcommand{\Memo}{\begin{MEMO}}
\newcommand{\eMe}{\end{MEMO}}

\newtheorem{ALGORITHM}[THM]{Algorithm}
\newcommand{\Algorithm}{\begin{ALGORITHM}}
\newcommand{\eAl}{\end{ALGORITHM}}

\usepackage[textsize=scriptsize,textwidth=2.5cm,shadow]{todonotes}

\usepackage{lineno}

\begin{document}


\title{A new approach to bipartite stable matching
optimization}

\author{Tam\'as Fleiner \thanks{Department of Computer Science and
Information Theory, Budapest University of Technology and Economics,
Magyar tud\'osok k\"or\'utja 2, Budapest, H-1117.  Research was
supported by ELKH-ELTE Egerv\'ary Research Group and the K143585 OTKA
Grant.  E-mail:  {\tt fleiner\char'100 cs.bme.hu} } \ \ \ \ \
{Andr\'as Frank \thanks{ELKH-ELTE Egerv\'ary Research Group,
Department of Operations Research, E\"otv\"os Lor\'and University
Budapest, P\'azm\'any P. s. 1/c, Budapest H-1117.  E-mail:  {\tt
andras.frank\char'100ttk.elte.hu } } \ \ \ \ \ {Tam\'as Kir\'aly
\thanks{ELKH-ELTE Egerv\'ary Research Group, Department of Operations
Research, E\"otv\"os Lor\'and University Budapest, P\'azm\'any P. s.
1/c, Budapest H-1117.  Research was supported by the Lend\"ulet
Programme of the Hungarian Academy of Sciences - grant number
LP2021-1-1/2021, by the Ministry of Innovatpon and Technology of
Hungary from the National Research, Development, and Innovation Fund -
grant ELTE TKP 2021-NKTA62 and grant  ADVANCED 150556.  E-mail:
{\tt tamas.kiraly\char'100 ttk.elte.hu }}} } }

\medskip

\maketitle

\medskip

\begin{abstract} As a common generalization of previously solved
optimization problems concerning bipartite stable matchings, we
describe a strongly polynomial network flow based algorithm for
computing $\ell$ disjoint stable matchings with minimum total cost.
The major observation behind the approach is that stable matchings, as
edge sets, can be represented as certain cuts of an associated
directed graph.  This allows us to use results on disjoint cuts
directly to answer questions about disjoint stable matchings.  We also
provide a construction that represents stable matchings as
maximum-size antichains in a partially ordered set (poset), which
enables us to apply the theorems of Dilworth, Mirsky, Greene and
Kleitman directly to stable matchings.  Another consequence of these
approaches is a min-max formula for the minimum number of stable
matchings covering all stable edges.  \end{abstract}

\noindent {\bf Keywords}:  \ stable matchings, packing and covering,
polynomial algorithms, network flows, posets, chains and antichains

\section{Introduction} \label{ahalintro}

By a bipartite preference system, we mean a bipartite graph
$G=(U,W;E)$ (with possible parallel edges) endowed with a (strict)
preference list of the edges (from better to worse) incident to $v$
for every node $v$ of $G$.  Sometimes we refer to the elements of $U$
as boys while the elements of $W$ are girls.

A matching $M$ of $G$ is called stable if it dominates every
non-matching edge $f=uw$ in the sense that $M$ has an element $e=u'w$
which is girl-better (at $w$) than $f$ or $M$ has an element $e=uw'$
which is boy-better (at $u$) than $f$.  The set of stable matchings
will be denoted by $\cal SM$ $={\cal SM}(G)$.  The starting result of
the area is the theorem of Gale and Shapley \cite{Gale+Shapley62}
stating that a bipartite preference system always admits a stable
matching, and, in addition, each stable matching covers the same
node-set, implying that they have the same cardinality.

Since the groundbreaking work of Gale and Shapley, extensive research has been done on finding efficient algorithms for more involved
optimization problems over the set of stable matchings.  In
particular, two fundamental approaches have been proposed for solving
the weighted stable matching problem in bipartite graphs.

The first approach relies on linear programming.  Polynomial-size
polyhedral descriptions of the stable matching polytope of bipartite
preference systems have been found by Vande Vate \cite{VandeVate} and
by Rothblum \cite{Rothblum92}.  Thus, general LP-solving techniques
can efficiently solve the weighted stable matching problem.  The
properties of the stable matching polytope were further explored by
Roth, Rothblum and Vande Vate \cite{RRV93} and by Teo and Sethuraman
\cite{TS98}, among others.

The second approach, presented in detail in the book of Gusfield and
Irving \cite{Gusfield+Irving}, is based on a one-to-one correspondence
between stable matchings and members of a certain ring-family.
Actually, they consider two different models.  Both give rise to a
network flow based algorithm for finding a maximum weight (or minimum
cost) stable matching, but the first one is conceptually simpler,
while the second one, relying on the concept of rotations, is
algorithmically more efficient.  Actually, this algorithm, developed
by Irving, Leather, and Gusfield \cite{ILG87} considers only the
special case when the cost-function reflects (or defines) the
preference list of persons.  It was later observed by several
researchers (see e.g., \cite{Dean+Munshi} and \cite{Kiraly+Pap}) that
the same technique works for general edge costs, as well, and yields a
strongly polynomial algorithm.

Here we introduce a new ring-family to model the structure of stable
matchings.  Like the ring-family of Gusfield and Irving, the members
of this one are also in a one-to-one correspondence with stable
matchings, where intersection and union correspond to the standard
meet and join operations on stable matchings.  The model describes a
direct correspondence between the set of stable matchings of a
bipartite graph and a set of certain $st$-cuts (defining the members
of the ring-family) of an associated digraph having only $\vert E_{\rm
st}\vert -\vert W\vert +2$ nodes (where $E_{\rm st}$
denotes the set of stable edges).  We remark that our model also avoids
the concept of rotations.  There is, however, a major difference between
the ring-family in the book of Gusfield and Irving
\cite{Gusfield+Irving} and the one in our approach.

The point is that a ring-family representation in itself is not enough
for solving packing and covering problems on stable matchings, because
it does not necessarily encode edge-disjointness.  To give an obvious
example, a preference system with two stable matchings can be
represented as a ring family on a single element, with the two sets of
the family being the emptyset and the one-element set.  However, this
representation provides no information on whether the two stable
matchings are edge-disjoint.  The approach in \cite{Gusfield+Irving}
does not address this problem, as it does not discuss packing and
covering problems.

The major advantage of our ring-family model is that, beyond handling
maximum weight stable matching problems, it also provides min-max
formulas and MFMC-based algorithms for various packing and covering
problems.  A basic packing problem of stable matchings aims at finding
a maximum number of (pairwise) disjoint stable matchings (or
equivalently, deciding if there are $\ell$ disjoint stable matchings).
Perhaps a bit surprisingly, this was considered and solved only
recently, by Ganesh et al.  \cite{GPNP21}.

Our present model makes it possible to manage a common generalization
of the weighted stable matching problem and the packing problem.  For
example, we describe an algorithm for finding $\ell$ disjoint cheapest
stable matchings.  In fact, we solve the even more general problem of
finding $\ell$ disjoint stable matchings with minimum total cost.  As
a consequence, an algorithm is described to determine the minimum
number of stable matchings covering all stable edges (the ones
occurring in some stable matching). Furthermore, based on our approach, we design a polynomial-time algorithm to find so-called level-fair stable matchings. This yields an efficient algorithm for various generalizations of previously studied fair stable matchings.

Beside the new ring-family model, we also introduce another approach.
Here the ground-set is the set $E_{\rm st}$ of stable edges, and we
define a certain partial order on $E_{st}$ induced by the reference
system on $G$ to get a poset $P_G$.  This poset has the specific
feature that all its inclusion-wise maximal antichains have the same
size, and these largest antichains are precisely the stable matchings
of $G$.  This poset model allows us to use packing and covering
results for antichains (like the ones of Dilworth, Mirsky, or Greene
and Kleitman) in order to solve packing and covering problems for
stable matchings.  For example, we describe min-max formulas and
strongly polynomial algorithms both for the minimum number of stable
matchings covering all stable edges, and for the maximum $w$-weight of
the union of $\ell$ stable matchings.  Based on this correspondence,
we describe a two-phase greedy algorithm for finding a minimum number
of stable matchings covering a lower-bound function $f$ on the set of
stable edges.

\subsection{Notions and notation}

Let ${\bf R, Q, Z} $ denote the set of real, rational, and integer
numbers, respectively.  When only non-negative values are allowed, we
use the notation ${\bf R_+, Q_+, Z_+} $. When $+\infty $ is also
allowed, we use the notation ${\bf \ol R_+, \ol Q_+, \ol Z_+} $. We
shall consider $+\infty $ as an integer.  For a function
$h:V\rightarrow {\bf R} $ and a subset $Z\subseteq V$, we use the
notation $\widetilde h(Z):= \sum [h(v):  v\in V]$. For a vector $x\in \mathbf{R}^n$, the vector $x^+$ is defined by $x^+_i=\max\{x_i,0\}$ ($i=1,\dots,n$).

For two elements $u$ and $v$ of a ground set $V$, a subset $Z\subset
V$ is a {\bf $v\ol u$-set} if $v\in Z\subseteq V-u$.  For a family
$\cal F$ of subsets, $\cup {\cal F}$ denotes the union of the members
of $\cal F$.  A subset $X\subseteq V$ is said to {\bf block} the
family $\cal F$ if $X$ intersects each member of $\cal F$.  Such a set
$X$ is also called a {\bf blocker} of $\cal F$, while an inclusionwise
minimal blocker is referred to as a {\bf minimal blocker}.

A set (collection) of distinct subsets of a ground set is called a
{\bf set-system,} while a collection of not necessarily distinct
subsets is a {\bf family} of sets.  A set-system $\cal R$ is called a
{\bf ring of sets} or a {\bf ring-set} (or just a ring)
if it is closed under the operations of intersection and union.  When
a subset is allowed to appear in more than one copies, we speak of a
{\bf ring-family}.

By adding the empty set and the ground-set, we obtain again a
ring-set, and thus we may apriori assume that $\{\emptyset
,V\}\subseteq {\cal R}$.  The sets $\emptyset $ and $V$ are the {\bf
trivial} members of $\cal R$ while the other members are {\bf
non-trivial}.  A ring-set is {\bf non-trivial} if it has a non-trivial
member.  The set of non-trivial members of a ring-set $\cal R$ will be
denoted by ${\cal R}'$.

Although a ring-set $\cal R$ may have an exponential number of members,
it can be encoded with the help of a function $C_{\cal R}:
V\rightarrow 2\sp V$, where $ C_{\cal R}(u) $ is the unique smallest
member of $\cal R$ \ containing $u$ (that is, $ C_{\cal R}(u) $ is the
intersection of all members of $\cal R$ containing $u$).

We call the function $C_{\cal R}$ the {\bf code} of $\cal R$.  It
consists of $\vert V\vert $ non-empty subsets of $V$.  A non-empty set
$Z\subseteq V$ is in $\cal R$ precisely if $C_{\cal R}(u)\subseteq Z$
holds for every element $u\in Z$ (that is, $Z= \cup (C_{\cal
R}(u):u\in Z)$).
If $v\in C_{\cal R}(u)-u$ for some $v\in V-u$, then the arc $uv$ will
be referred to as a {\bf code-arc} of $\cal R$.  The digraph $D_{\cal
R}$ on $V$ formed by the code-arcs is called the {\bf code-digraph} of
$\cal R$.  A subset $Z$ of $V$ is a member of $\cal R$ if and only if
no arc of the code-digraph leaves $Z$.

From an algorithmic point of view, when we say that we are given a
ring-set $\cal R$, it means that $\cal R$ is given by its code or
code-digraph.  Another natural subroutine to encode $\cal R$ tells for
any ordered pair of nodes in $V$ whether or not there is a $u\ol
v$-member of $\cal R$.  Obviously $v$ is in $C_{\cal R}(u)$ if and
only if the answer is no, and hence the two descriptions of $\cal R$
are equivalent.

Let $D=(V,A)$ be a loopless digraph, $s\sp *\in V$ a specified
source-node with no entering arcs, and $t\sp * \in V$ a specified
sink-node with no leaving arcs.  Let ${\cal R}\sp *$ denote the
ring-set consisting of the empty set, the ground-set $V$, and all the
$s\sp *\ol{t\sp *}$-subsets of $V$.

For a subset $ X \subseteq V$, let OUT$_D(X)$ denote the set of arcs
leaving $X$.  By an {\bf $s\sp *t\sp *$-cut} of $D$ we mean the set of
arcs leaving an $s\sp *\ol{t\sp *}$-subset $Z\subset V$, where $Z$ is
called the {\bf out-shore} of the cut.

In the standard Max-flow Min-cut (MFMC) problem, we are given a
capacity-function $g:A\rightarrow \ol{{\bf R}}_+$ (serving as an upper
bound) on the set of arcs of $D$.  We say that $g$ is integer-valued
if its finite values are integers.

Let the {\bf out-capacity} $\delta _g(Z)$ of a subset $Z\subseteq V$
of nodes be the $g$-sum of the arcs leaving $Z$ while $\varrho
_g(Z):=\delta _g(V-Z)$.  For every flow $x\geq 0$ in $D$, we have
$\delta _x(s\sp *) = \varrho _x(t\sp *)$ and this common value is the
{\bf amount} of $x$.  A simple property of flows $x$ is that $\delta
_x(Z)-\varrho _x(Z)=\delta _x(s\sp *)$ holds for every $s\sp *
\ol t\sp *$-set $Z$.  A flow $x$ is {\bf $g$-feasible} or just {\bf
feasible} if $x\leq g$, and this is why sometimes we refer to $g$ as a
capacity-function.

The primal MFMC problem aims at finding a feasible flow of maximum
flow amount while the dual problem is about finding $s\sp *t\sp *$-cut
with minimal $g$-value or equivalently finding an $s\sp *\ol{t\sp
*}$-set $Z\subset V$ of minimum $\delta _g(Z)$ value.

The classic MFMC theorem of Ford+Fulkerson states that $\max = \min$,
and if $g$ is integer-valued, then the maximum flow may be chosen
integer-valued.  By the algorithms of Edmonds and Karp, Dinits, or
Goldberg and Tarjan, both a maximum flow and a minimum cut can be
computed in strongly polynomial time (when $g$ is rational- or
integer-valued).  For a general overview of these algorithms, see the
book of Schrijver \cite{Schrijverbook}.

\subsection{Stable matchings}

The books of Gusfield and Irving \cite{Gusfield+Irving} and Manlove
\cite{Manlove-book} provide a rich overview of definitions and results
on stable matchings.  Here we recall some of those which are important
in the present work.

An edge of $G$ is called {\bf stable} if it belongs to a stable
matching.  Let $E_{\rm st}$ denote the set of stable edges (that is,
the union of all stable matchings).  An edge $e$ is {\bf marginal} if
its removal does not affect the set of stable edges.  The removal of a
marginal edge $f$ may result in a new marginal edge, and it can also 
happen that an originally marginal edge $h$ ceases to be marginal, that
is, $G-f-h$ may have a new stable matching.

An easily provable basic observation is that if we consider the best
edge $e=st$ incident to a node $s$, then any edge incident to $t$
which is worse than $e$ at $t$ is marginal, and hence its deletion
does not affect the set of stable edges.  Therefore, we can delete all
the edges incident to $t$ that are worse than $e$.  By going through
all nodes $s$ of $G$ and carrying out this edge-deletion process, and
finally deleting the arising singleton nodes, we obtain a graph in
which the set of stable matchings is the same as in the starting
graph.  (Note that this procedure is essentially a proof of the
theorem of Gale and Shapley stating that there always exists a stable
matching and that the set of nodes covered by a stable matching is the
same for each stable matching.)  Furthermore, the set of girl-best
edges is a stable matching (which coincides with the set of boy-worst
edges) and, symmetrically, the set of boy-best edges is a stable
matching (which coincides with the set of girl-worst edges).

Therefore, we may assume henceforth that the original graph $G=(U,W;E)$
itself has the property that each stable matching is a perfect
matching, and both the set of girl-best edges and the set of boy-best
edges form a stable matching.  Let $n := \vert W\vert $ ($=\vert
U\vert $).

Another fundamental property (see Lemma 1.3.1 in
\cite{Gusfield+Irving}) is that if $M$ and $N$ are two (distinct)
stable matchings, then the set of girl-best edges in $M\cup N$ is a
stable matching, denoted by $M\wedge N$, as well as the set of
girl-worst edges, denoted by $M\vee N$.  Moreover, $M\wedge N$ is the
set of boy-worst edges, while $M\vee N$ is the set of boy-best edges.
These two operations define a distributive lattice on the set of
stable matchings.  Actually, the following generalization is also
valid.

\Claim \label{unioban-stabil} If $E'$ is the union of an arbitrary set
of stable matchings, then the set of girl-best edges in $E'$ is a
stable matching and so is the set of the girl-worst edges, as well.
The set of girl-best edges coincide with the set of boy-worst edges,
and, symmetrically, the set of girl-worst edges coincide with the set
of boy-best edges.  \FBOX \eCl

By applying this claim to the set of stable matchings containing a
stable edge $e$, one obtains that in the union of stable matchings
containing $e$ there is a unique girl-best stable matching containing
$e$, denoted by $M_e$, which coincides with the boy-worst stable
matching containing $e$.

We can conclude that the arguments and algorithms concerning the set
of all stable matchings of $G$ can be extended to the set of stable
matchings containing a given stable edge $e$.  For example, for the
algorithms to be developed below, we may assume that the (unique)
girl-best stable matching $M_e$ containing $e$ is available for each
stable edge $e$ of $G$.  Claim \ref{unioban-stabil} immediately
implies the following.

\Claim \label{Mis} For any stable matching $M=\{e_1,\dots e_n\}$ of
$G$, one has $M= M_{e_1}\vee \cdots \vee M_{e_n}$.  \FBOX \eCl

We call a matching of $G$ {\bf stable extendible} or just {\bf
extendible} if it is a subset of a stable matching.

\Claim \label{Nextend} A matching $M$ of $G$ is stable extendible if
and only if $M\cup M'$ is a matching for any (and hence for each)
stable matching $M'$ of $G'$, where $G'$ denotes the graph arising
from $G-M$ by deleting all edges dominated by $M$.  In particular,
with the help of a single application of the Gale+Shapley algorithm
one can decide if a matching $M$ of $G$ is stable extendible or not,
and if so, the algorithm provides the unique girl-best stable matching
including $M$.  \eCl

\Proof If $M\cup M'$ is a matching for a stable matching $M'$ of $G'$,
then it is clearly a stable matching of $G$, that is, $M$ is
extendible.  Conversely, if $M$ can be extended to a stable matching
$M\cup M'$ of $G$, then $M'$ must be a stable matching of $G'$, since
if $G'$ had an edge $e$ not dominated by $M'$, then $e$ is dominated
by an element of $M$ but such an edge $e$ is not in $G'$.  \FBOX

\section{Optimization over ring-sets defined on a digraph}

In this preparatory section, we do not discuss stable matchings at
all.  Instead, we investigate some flow and tension problems
concerning ring-sets, which shall be used in forthcoming sections to
solve various optimization problems on stable matchings.

\subsection{Largest $h$-independent packing of ring-set members}

Let $D=(V,A)$ be a loopless digraph with a source-node $s\sp *$ and a
sink-node $t\sp *$.  We assume that every node is reachable from $s\sp
*$.  Let $h :  A\rightarrow {\bf Z}_+$ be a non-negative
integer-valued function on the arc-set of $D$.  We call a family of
subsets of nodes {\bf $h$-independent} if every arc $a$ of $D$ leaves
at most $h(a)$ members of the family.  In the special case of $h\equiv
1$, we speak of {\bf arc-independence}.  A subset $L\subseteq A$ of
arcs {\bf out-covers} a subset $X\subset V$ of nodes if it contains at
least one arc leaving $X$, that is, if $\delta _L(X)\geq 1$.  For a
set-system $\cal F$, we say that $L$ {\bf out-covers} $\cal F$ if $L$
out-covers each member of $\cal F$.

Let ${\cal R}_0$ \ $(\subseteq {\cal R}\sp *)$ be a non-trivial
ring-set (given by its code) on the node-set of digraph $D=(V,A)$, and
suppose that each non-trivial member of ${\cal R}_0$ is an $s\sp *\ol
{t\sp *}$-set.  Recall that ${\cal R}_0'$ denotes the set of non-trivial
members of ${\cal R}_0$.  Let $\nu _h$ denote the maximum number of
(not necessarily distinct) $h$-independent members of ${\cal R}'_0$,
and let

$$ \hbox{ $\tau _h := \min \{\widetilde h(L):  \ L\subseteq A, \ L$
out-covers ${\cal R}'_0\}$.  }\ $$

\THEOREM \label{pakol} For digraph $D=(V,A)$, function $h$, and
ring-set ${\cal R}_0$, we have $\nu _h = \tau _h$.  By a single
application of Dijkstra's shortest path algorithm, both a
$h$-independent family of $\nu _h$ (not necessarily distinct) members of ${\cal R}_0$ and a
subset $L\subseteq A$ of $\tau _h$ ($=\nu _h)$ arcs out-covering
${\cal R}'_0$ can be computed.  In addition, a maximum $h$-independent
family may be chosen so as to form a chain.  \eT

\Proof As the inequality $\max \leq \min$ is straightforward, we
concentrate on the reverse inequality.  For every node $u\in V$ and
node $v\in C_{\tiny {\rm \cal R}_0}(u)$, add the code-arc $uv$ to $D$,
and let $D'=(V,A')$ denote the extended digraph.  We shall refer to
the members of $A$ as original arcs.  By its definition, no code-arc
leaves any set $Z\in {\cal R}_0$.  Define a cost-function $c'
:A'\rightarrow {\bf Z} _+$, as follows.

\eq c'(a) := \begin{cases} h(a) & \ \ \hbox{if}\ \ \ e\in A \\ 0 & \
\ \hbox{if}\ \ \ a\in A'-A.  \end{cases} \eeq

By a single application of Dijkstra's algorithm, we can compute an
$s\sp *$-rooted spanning arborescence of $D'$ in which the (unique)
$s\sp *v$-path is a $c'$-cheapest $s\sp *v$-path of $D'$ for all nodes
$v$.  Let $\pi '(v)$ denote the $c'$-cost of this path.

Since there is a code-arc from $t\sp *$ to every other node, and the
$c'$-cost of every code-arc is $0$, we have $\pi '(s\sp *)=0 \leq \pi
'(v) \leq \pi' (t\sp *)$
for every node $v$.  If $\pi' (t\sp *)=0$,
then
$D'$ includes an $s\sp *t\sp *$-path $P$ of 0 cost.  Let $P'$ denote
the subset of original arcs of $P$.  Since no code-arc leaves any
member of ${\cal R}_0$, $P'$ out-covers ${\cal R}_0$.  As the total
$h$-cost of $P'$ is $0$, we obtain that $\nu _h \leq \tau _h = 0 \leq
\nu _h$, that is, $\nu _h=\tau _h$.

Therefore, we may assume that $\pi '(t\sp *) >0$.  Let $q$ \ ($q\geq
1$) denote the number of distinct positive values of $\pi '$ and let
$(0=) \ \mu _0 < \mu _1 < \cdots < \mu _q \ ( = \pi '(t\sp *))$ denote
the distinct values of $\pi '$.

A well-known property is that the (integer-valued) function $\pi '$ is
a feasible potential in the sense that that $\Delta _{\pi '}(a) \leq
c'(a)$ holds for every arc $a=uv$ of $D'$ where $\Delta _{\pi '}(a) :=
\pi '(v)-\pi '(u)$ denotes the {\bf potential-drop} induced by $\pi
'$.  Moreover, an $s\sp *t\sp *$-path is $c'$-cheapest if and only if
it consists of tight arcs (where tight means that $\Delta _{\pi '}(a)
= c'(a)$).

Consider the level sets $L_0, L_1, \dots , L_{q-1} \ (\subseteq V-t\sp
*)$ defined by $L_i :=\{v\in V:  \pi '(v)=\mu _i \}$, along with the
induced chain ${\cal C}:= \{V_0\subset V_1\subset \cdots \subset
V_{q-1}\}$ where $V_i:= L_0\cup \cdots \cup L_i$.  Since no set $V_i$
is left by any arc with $0$ $c'$-cost, it follows that each $V_i$ is
in \ ${\cal R}_0$.

Let ${\cal C}'$ denote the chain of sets in which each set $V_i$
occurs $\mu _{i+1}-\mu _i$ times ($i=0,\dots ,q-1$). We emphasize that ${\cal C}'$ is a family of sets.
Then $\vert
{\cal C}'\vert = \mu _q$.  Let $P$ be $c'$-cheapest $s\sp *t\sp
*$-path of $D'$ whose $c'$-cost is $\mu _q$.  It follows from the
feasibility of $\pi '$ that $P$ leaves each set $V_i$ exactly once
(that is, $P$ does not enter $V_i$).  Furthermore, the unique arc of
$P$ leaving $V_i$ is an original arc whose $c'$-cost (that is, its
$h$-value) is $\mu _{i+1}-\mu _i$.

It follows from these that the chain ${\cal C}'$ consists of $\mu _q$
members of ${\cal R}'_0$ and these sets form an $h$-independent family,
from which $\nu _h\geq \mu _q$.  On the other hand, the original arcs
of $P$ out-cover ${\cal R}'_0$, and the sum $\widetilde {c'}(P)$ of
$h$-values of these arcs is $\mu _q$.  Therefore we have $\tau _h\leq
\mu _q \leq \nu _h \leq \tau _h$, and hence $\nu _h=\tau _h$ follows.
\FBOX

\subsubsection{A two-phase greedy algorithm for the special case
$h\equiv 1$} \label{direkt}

In the special case $h\equiv 1$, Theorem \ref{pakol} can be
reformulated in the following simpler form.

\Corollary \label{pakolb} In digraph $D=(V,A)$, let ${\cal R}_0$ \
$(\subseteq {\cal R}\sp *)$ be a non-trivial ring-set (given by its
code).  Then the maximum number $\nu _1$ of the arc-independent
members of ${\cal R}_0$ is equal to the minimum number $\tau _1$ of
arcs out-covering ${\cal R}_0$.  The optimal arc-independent system
may be chosen to be a chain.  With the help of a two-phase greedy
algorithm, both a largest arc-independent chain and a smallest arc-set
out-covering ${\cal R}_0$ can be computed in polynomial time.  \eCo

\Proof The Dijkstra algorithm occurring in the proof of Theorem
\ref{pakol}, when applied to the case $h\equiv 1$, is concerned with a
special $(0,1)$-valued cost-function.  In this case, the Dijkstra
algorithm (for computing the largest arc-independent chain ${\cal C} =
\{V_0\subset \cdots \subset V_{q-1}\} \ \subseteq \ {\cal R}'_0$ along
with an $s\sp *t\sp *$-path consisting of a minimum number of original
arcs) can be replaced by the following two-phase greedy algorithm.

Phase 1 consists of subsequent steps for $i=1,2,\dots $. In Step 1, let
$V_1$ be the (unique) smallest member of ${\cal R}_0$ containing $s\sp
*$.  Since ${\cal R}_0$ is non-trivial, $t\sp *\not \in V_0$.  In Step
$i+1$ of the first phase ($i=1,2,\dots $), consider the set $V_{i}$
computed in the previous step.  Let $V_{i}'$ denote the set of nodes
consisting of $V_{i}$ and the heads of original arcs leaving $V_{i}$.

If there is no member of ${\cal R}_0$ including $V'_i$ but not
containing $t\sp *$, then we define $q:=i$ and Phase 1 terminates by
outputting the arc-independent chain $V_1\subset \cdots \subset V_{q}$
consisting of members of ${\cal R}_0$.  If ${\cal R}_0$ has a member
including $V'_i$ but not containing $t\sp *$, then let $V_{i+1}$ be
the unique smallest such member.  (This is nothing but the set of
nodes reachable in $D'$ from $V'_i$.)

In Phase 2, by starting at node $t\sp *$ and stepping back one-by-one,
we can build up in a greedy way a reverse dipath from $t\sp *$ to
$s\sp *$ which consists of exactly $q$ original arcs and at most $q$
code-arcs.  Let $P$ denote the corresponding $s\sp *t\sp *$-dipath in
$D'$.  By the construction, $P$ leaves each $V_i$ ($i=1,\dots ,q$)
exactly once along an original arc and all other arcs of $P$ are
code-arcs.  Moreover, since no member of ${\cal R}_0$ is left by a
code-arc, the $q$ original arcs of $P$ out-cover every member of
${\cal R}_0$.  Therefore, we have $\tau _1 \leq q \leq \nu _1\leq \tau
_1$ and hence $\tau _1 =q = \nu _1 =\tau _1$.  \FBOX

\subsection{Minimizing the out-capacity $\delta _g(Z)$ over the
members of a ring-set} \label{ming}

Let $D=(V,A)$ be a digraph with a source-node $s\sp *$ and a sink-node
$t\sp *$.  Let $g:A\rightarrow \ol{{\bf Z}}_+$ be an upper bound or
capacity function on the arc-set $A$, where \ $\ol{{\bf Z}}_+ = {\bf
Z}_+ \cup \{+\infty \}$.  As is well-known, there are strongly
polynomial algorithms \cite{Dinits, Edmonds-Karp} for computing a
minimum $g$-capacity $s\sp *t\sp *$-cut, or equivalently, an $s\sp
*\ol t\sp *$-set $Z$ that minimizes $\delta _g(Z)$.  Our present goal
is to show how such an algorithm (working with augmenting paths) can
be used to compute a member $Z$ of a non-trivial ring-set ${\cal R}_0
\subseteq {\cal R}\sp *$ (given by its code) for which $\delta _g(Z)$
is as small as possible.  Actually, it turns out that the minimizer
sets also form a ring-set, and we not only compute a single minimizer
but the code of this ring-set.  This will be an important tool in
stable matching applications we discuss later.

Recall that for a ring-set ${\cal R}_0$, ${\cal R}'_0$ denotes the set
of non-trivial members of ${\cal R}_0$.

\Lemma \label{code} Let ${\cal R}_0 \subseteq {\cal R}\sp *$ be a
non-trivial ring-set given by its code.  Suppose that $\delta
_g(Z)\geq 1$ holds for each member $Z$ of ${\cal R}_0$ (which is an
$s\sp * \ol {t\sp *}$-set).  Let $$ \hbox{ $\gamma _1 := \min \{\delta
_g(Z):  Z\in {\cal R}'_0 \} $ \ and \ ${\cal R}_1 :=\{Z\in {\cal R}_0:
\delta _g(Z) = \gamma _1\} \cup \{\emptyset ,V\}$.  }\ $$ Then ${\cal
R}_1$ is a ring-set whose code can be computed in strongly polynomial
time by an MFMC algorithm.  \eL

\Proof If $\gamma _1=\infty $, then $\delta _g(X)=\infty $ holds for
every non-trivial member $X$ of ${\cal R}_0$.  In this case, ${\cal
R}_1 = {\cal R}_0$, and hence the code of ${\cal R}_1$ is ab ovo
available.

Therefore, we can assume that $\gamma _1$ is finite. First we observe that system ${\cal R}_1$ is a ring-set. As $\delta _g$ is a submodular set-function, for sets $X,Y\in {\cal R}'_1$ we have $$\gamma _1+\gamma _1 =
\delta _g(X) + \delta _g(Y) \geq \delta _g(X\cap Y) + \delta _g(X\cup
Y) \geq \gamma _1+\gamma _1,$$ from which we have $\delta _g(X\cap
Y)=\gamma _1$ and $\delta _g(X\cup Y) =\gamma _1$.

To prove the second part, we extend $D$ by adding the code-arcs of
${\cal R}_0$.  We also extend $g$ to the code-arcs $a'$ by letting
$g(a'):=+\infty $.
Let $D'=(V,A')$ denote the extended digraph, and $g'$ the extended
capacity-function.

Since no code-arc leaves any member $Z$ of ${\cal R}_0$, we have
$\delta _g(Z) = \delta _{g'}(Z)$.  Furthermore, $\delta _{g'}(X) =
+\infty $ holds for every $s\sp *\ol{t\sp *}$-set $X \not \in {\cal
R}_0$, so it follows that the $\delta _{g'}$-minimizer $s\sp *\ol{t\sp
*}$-sets of $D'$ are the members of \ ${\cal R}'_1$ (that is, the
$\delta _g$-minimizer members of ${\cal R}'_0$).  \

Let $x$ be a $g'$-feasible flow in $D'$ with maximum flow amount (in
short, a maximum flow) whose flow amount by the MFMC theorem is
$\gamma _1$.  Consider the standard auxiliary digraph $D''$
(introduced in the Ford-Fulkerson MFMC algorithm) associated with flow
$x$ (in which $uv$ is an arc if $uv\in A'$ and $x(uv)<g'(uv)$, or if
$vu\in A'$ and $x(vu)>0$).

Now ${\cal R}_1 = \{Z\in {\cal R}\sp * :  \delta _{D''}(Z) = 0\}$, and
hence $C_{\tiny {\rm \cal R}_1}(u)$ is nothing but the set of nodes
reachable from $u$ in $D''$.  (which always contains $s\sp *$ by the
definition of $D''$).  Therefore, the code of ${\cal R}_1$ can be
computed by $\vert V\vert $ applications of a path-finding (or
reachability) subroutine.  \FBOX

\medskip The lemma can easily be extended to the case when not only
one single function $g$ is given on $A$ but more.

\Corollary \label{code1} Let ${\cal R}_0$ \ $( \subseteq {\cal R}\sp
*)$ be a non-trivial ring-set given by its code, and let
$g_1,g_2,\dots ,g_k$ be non-negative integer-valued functions on the
arc-set of digraph $D$.  Let $$ \hbox{ $\gamma _1 := \min \{\delta
_{g_1}(Z):  Z\in {\cal R}'_0 \}$ and ${\cal R}_{1} :=\{Z\in {\cal
R}'_0 :  \delta _{g_1}(Z) = \gamma _1\} \cup \{\emptyset ,V\}, $ }\ $$
and for $i=2,\dots ,k$ let $$ \hbox{ $\gamma _i := \min \{\delta
_{g_i}(Z):  Z\in {\cal R}'_{i-1} \}$ and ${\cal R}_{i} :=\{Z\in {\cal
R}'_{i-1} :  \delta _{g_i}(Z) = \gamma _i\} \cup \{\emptyset ,V\}.  $
}\ $$ Then each of the set-systems ${\cal R}\sp * \supseteq {\cal R}_1
\supseteq \cdots \supseteq {\cal R}_k$ is a ring-set whose codes can
be computed in strongly polynomial time with an MFMC algorithm.  In
particular, the non-trivial members of ${\cal R}_k$ are exactly those
non-trivial sets $Z\in {\cal R}_0$ for which $\delta _{g_1}(Z)$ is
minimum, and     this, $\delta _{g_2}(Z)$ is minimum, and within
this $\delta _{g_3}(Z)$ is minimum, and so on.  \eCo

\Proof By applying Lemma \ref{code} separately $k$ times in a sequence
to the ring-sets ${\cal R}_0,\dots , {\cal R}_{k-1}$, the statement
follows immediately.  \FBOX

\section{Associating a digraph with the preference system on $G$}
\label{assoc}

Let $G=(U,W;E)$ be a bipartite graph endowed with a preference system,
and let $E_{\rm st}$ denote the set of stable edges.  We assume
throughout that each stable matching is a perfect matching.  Let
$n:=\vert U\vert =\vert W\vert $. The goal of this section is to
associate a digraph $D$ with $G$ along with a ring-set ${\cal R}_D$ on
$D$ in such a way that there will be a simple one-to-one correspodence
between the stable matchings of $G$ and the cuts belonging to the
members of ${\cal R}_D$ (that is, the arc-sets OUT$_D(X)$ for $X\in
{\cal R}_D$).

We emphasize already here that the role of girls and boys in this
definition is asymmetric, since only the girl preferences play a direct
role.  (Of course, the boy preferences are implicitly involved in the
set of stable edges.)

Let us define a digraph $D=(V,A)$, as follows.  $D$ will have two
types of arcs:  stable and dummy.  Let $s\sp *\in V$ be a source-node
and $t\sp *\in V$ a sink-node of $D$.  With every girl $w\in W$, we
associate a (one-way) $s\sp *t\sp *$-path $P_w$ in $D$ which will be
referred to as a {\bf girl-path} of $D$.  The arcs of $P_w$ correspond
to the stable edges of $G$ incident to $w$, and they follow each other
in the (girl) preference order of the stable edges incident to $w$.
In particular, the first arc of $P_D$ (whose tail is $s\sp *$)
corresponds to the girl-best edge at $w$, while the last arc of $P_w$
(whose head is $t\sp *$) corresponds to the girl-worst edge at $w$.
The girl-paths are internally disjoint, and we shall refer to the arcs
of girl-paths as {\bf stable arcs} of $D$.  Therefore, $D$ has $\vert
E_{\rm st}\vert -\vert W\vert +2$ nodes.  The stable arc of $D$
assigned to a stable edge $e\in E$ of $G$ will be denoted by
$a_e=t_eh_e$, where $t_e$ is the tail of $a_e$ and $h_e$ is the head
of $a_e$.

Before defining the dummy arcs of $D$, consider a stable matching
$M=\{e_1,\dots ,e_n\}$ of $G$, and let $w_i$ denote the end-node of
$e_i$ in $W$.  Let $A_M:=\{a_{e_1},\dots ,a_{e_n}\}$ denote the set of
stable arcs of $D$ corresponding to the elements of $M$, and let
$L(M)$ denote the set of those nodes of $D$ which are on the subpath
of a girl-path $P_{w_i}$ starting at $s\sp *$ and ending at $t(a_i)$
for some $i=1,\dots ,n.  $

For every stable edge $e$ of $G$, consider the corresponding arc
$a_e=t_eh_e$ of $D$.  Let $t_ev$ be a {\bf dummy arc} of $D$ for each
node $v\in L(M_e)$.  In particular, this means that there is a dummy
arc from $t_e$ to every node preceding $t_e$ in the girl-path
containing $a_e$. The following claims directly follow from the previously discussed properties of stable matchings and the above definitions.

\Claim \label{ } A stable matching $N$ is girl-better than another
stable matching $M$ if and only if $L(N)\subseteq L(M)$.  \FBOX \eCl

\Claim Let $M$ be a stable matching and $e$ a stable edge which is
either in $M$ or girl-better than some member of $M$.  Then $L(M_e)
\subseteq L(M)$.  \FBOX \eCl

\Claim If $M$ and $N$ are stable matchings of $G$, then \eq L(M\wedge
N) = L(M) \cap L(N) \ \ \hbox{and}\ \ \ L(M\vee N) = L(M) \cup L(N).
\label{(teto-talp)} \eeq \FBOX \eCl

Let \eq \hbox{ $ {\cal R}_D :  = \{Z\subset V:$ an $s\sp *\ol{t\sp
*}$-set with no leaving dummy arc$\}$.}\ \label{(RD)} \eeq

\noindent The set-system ${\cal R}_D$ is is a ring-set.  By the
definition of dummy arcs, every girl-path leaves a member $Z$ of
${\cal R}_D$ exactly once, that is, $\delta _D(Z) =n$.  We call this
$n$-element set of arcs leaving $Z$ a {\bf stable $s\sp *t\sp *$-cut}
of $D$.

\subsection{Stable matchings of $G$ versus stable $s\sp *t\sp *$-cuts
of $D$}

The clue to our suggested solution of various optimization problems
concerning bipartite stable matchings is that there is a natural
one-to-one correspondence between the stable matchings of $G$ and the
stable $s\sp *t\sp *$-cuts of the digraph $D$ associated with $G$.
This is formulated in the next lemma.

\Lemma \label{corres} For every set $Z\in {\cal R}_D$, the $n$ stable
arcs of $D$ leaving $Z$ correspond to the $n$ elements of a stable
matching $M$ of $G$ for which $L(M) =Z$ and $A_M=${\rm OUT}$_D(Z)$.

Conversely, for every stable matching $M$ of $G$, the set $Z:=L(M)$ is
in ${\cal R}_D$, and the $n$ stable arcs of $D$ corresponding to the
$n$ elements of $M$ are the arcs leaving $Z$, that is, $A_M=${\rm
OUT}$_D(Z)$.  \eL

\Proof To prove the first part, consider the set OUT$_D(Z):=
\{a_1,a_2,\dots ,a_n\}$ of arcs leaving $Z$ (where $a_i$ is a member
of the girl-path $P_{w_i}$).

Let $e_1,\dots ,e_n$ denote the stable edges of $G$ corresponding to
the arcs $a_1,\dots ,a_i$, that is, $a_i= a_{e_i}$.  For each
$i=1,\dots ,n$, consider the girl-best stable matching $M_i:=M(e_i)$
containing $e_i$.

We claim that $L(M_i)\subseteq Z$, since if there were a node $v$ in
$L(M_i)-Z$, then the the dummy arc $t_{a_i}v$ would leave $Z$,
contradicting the property that no dummy arc leaves $Z$.  From these
it follows that $Z= \cup \{L(M_i):  i=1,\dots ,n\}$.

Consider now the stable matching $M':  = M_1\vee \cdots \vee M_n$.  It
follows from the second half of observation \eref{(teto-talp)} that
$L(M')= L(M_1)\cup \cdots \cup L(M_n) = Z$, and hence $A_{M'} =$
OUT$_D(Z) = A_M$, that is $M=M'$, from which the first half of the
lemma follows.

To prove the second half, let $M=\{e_1,\dots ,e_n\}$ be a stable
matching of $G$, let $Z:=L(M)$ and $A_M:= \{a_1,\dots ,a_n\}$.  By
Claim \ref{Mis}, we have $M= M_{e_1}\vee \cdots \vee M_{e_n}$.  It
follows from the second half of observation \eref{(teto-talp)} that
$Z=L(M)= L( M_{e_1}) \cup \cdots \cup L(M_{e_n})$.  Therefore no dummy
arc leaves $Z$, that is, OUT$_D(Z)= A_M$.  \FBOX

\medskip

Lemma \ref{corres} implies the following.

\THEOREM \label{GDcorres} \ Let $M$ be a stable matching of $G$, and
let $Z$ be a non-trivial member of ring-set ${\cal R}_D$.  Then \eq
\hbox{ $Z=L(M)$ \ holds if and only if \ {\rm OUT}$_D(Z)= A_M$.  }\
\label{(GDcorres)} \eeq The two equalities in \eref{(GDcorres)}
determine a one-to-one correspondence between the stable matchings $M$
of $G$ and the non-trivial members $Z$ of ring-set ${\cal R}_D$.
\FBOX \eT

Call a family of (not necessarily distinct) stable matchings
{\bf $h$-independent} if every stable edge $e$ belongs to at most
$h(e)$ stable matchings.

If we apply Theorem \ref{pakol} to the digraph associated with a
bipartite preference system, then Lemma \ref{corres} implies the
following.

\Corollary \label{gindep} Let $h$ be a non-negative integer-valued
function on the edge-set of a bipartite graph $G$ endowed with a
preference system.  The maximum number of $h$-independent stable
matchings
is equal to the minimum $h$-value of a blocker of stable
matchings.  In particular, the maximum number of disjoint stable
matchings is equal to the minimum number of edges blocking all stable
matchings.  Furthermore, by a single application of Dijkstra's
shortest path
algorithm, both a largest packing of $h$-independent
family of stable matchings and a minimum $h$-cost blocker of stable
matchings can be computed in strongly polynomial time.
\FBOX \eCo

We hasten to emphasize that in Theorem \ref{mincostpak2} we shall show
that the same approach works in the more complex situation when the goal
is finding a maximum number of $h$-independent $c$-cheapest stable
matchings where $c$ is a non-negative cost-function on the set of
stable edges.

\section{Cheapest stable matchings } \label{olcsost}

Let $c$ be an integer-valued cost-function on the set of stable edges
of $G$.  Rothblum \cite{Rothblum92} provided a particularly simple
polyhedral description of the polytope of stable matchings that uses
$(0,1)$-inequalities and the number of these inequalities is only
$O(n\sp 2)$.  Therefore the general purpose linear programming
algorithm of Tardos \cite{Tardos86} to solve combinatorial linear
programs in strongly polynomial time can be applied to compute a
$c$-cheapest stable matching.

By relying on the fundamental concept of rotations, a cheapest stable
matching can also be computed with the help of a standard network flow
subroutine, see the books of Gusfield and Irving
\cite{Gusfield+Irving} and of Manlove \cite{Manlove-book}.

The first goal of the present section is to describe a direct
algorithm to compute a cheapest stable matching in strongly polynomial
time that uses only network flows and does not need the concept of
rotations.  The second goal is to develop an algorithm for packing
$c$-cheapest stable matchings.

\subsection{How to find a cheapest stable matching}

In this section, we show how a $c$-cheapest stable matching of $G$ can
be computed with the help of a single MFMC algorithm that finds a
minimum capacity $s\sp *t\sp *$-cut of the digraph $D=(V,A)$
associated with the preference system on $G$ as described in Section
\ref{assoc}.  Even more, we can compute the code of the ring-set
defined by all cheapest stable matchings.

Since each stable matching has the same cardinality, one can shift $c$
by a constant, and hence it can be assumed that $c$ is non-negative
(even that $c$ is everywhere positive).  Let us define the capacity
function $g_c$ on $A$ as follows.

\eq \hbox{ $g_c(a)$ }\ := \begin{cases} c(e) & \ \ \hbox{if $a=a_e$ is
the stable arc of $D$ associated with $e\in E_{\rm st}$}\ \\ +\infty
& \ \ \hbox{if $a$ is a dummy arc of $D$.}\ \end{cases} \label{(gc)}
\eeq

Obviously $$ \hbox{ ${\cal R}_D=\{Z\subseteq V-t\sp *:  s\sp *\in Z, \
\delta _{g_c}(Z)$ is finite$\}$.  }\ $$

Theorem \ref{GDcorres} immediately gives rise to the following.

\THEOREM \label{cgcorres} \ For the correspondence described in
Theorem \ref{GDcorres} between the stable matchings $M$ of $G$ and the
non-trivial members $Z$ of ring-set ${\cal R}_D$, one has

$$ \hbox{ $\delta _{g_c}(Z) = \widetilde c(M)$.  }\ $$

\noindent This is a one-to-one correspondence between the $c$-cheapest
stable matchings $M$ of $G$ and those non-trivial members $Z$ of
ring-set ${\cal R}_D$ which minimize $\delta _{g_c}(Z)$.
Consequently, finding a $c$-cheapest stable matching of $G$ can be
done by computing an $s\sp *t\sp *$-cut of $D$ with minimum
$g_c$-capacity, which is doable by a strongly polynomial MFMC
subroutine.  \FBOX \eT

\medskip

\Corollary The set of cheapest stable matchings of $G$ is closed under
the operations meet $\wedge $ and join $\vee $, and hence there exists
a (unique) girl-best cheapest stable matching among all cheapest
stable matchings (which is the boy-worst cheapest matching).  The
cheapest stable matching provided by the algorithm mentioned in
Theorem \ref{cgcorres} provides this girl-best cheapest stable
matching.  \eCo

\Proof As is well known, the shores of minimum capacity $s\sp *t\sp *$
cuts containing $s\sp *$ form a ring-set ${\cal R}_{\rm min}$ \
($\subseteq {\cal R}_D$).  This implies the first part of the corollary
via Lemma \ref{corres}.  The MFMC algorithms of Edmonds and Karp or
the one by Dinits computes the unique smallest member of ${\cal
R}_{\rm min}$, from which the second part also follows.  \FBOX

\medskip

By applying Lemma \ref{code} to ring-set ${\cal R}_0:  = {\cal R}_D$,
we obtain the following.

\Corollary \label{mincost} Let ${\cal R}_1$ denote the set-system
consisting of the $\delta _{g_c}$-minimizer members of ring-set ${\cal
R}_1$ is a ring-set whose code can be computed in strongly polynomial
time.  The correspondence in \eref{(GDcorres)} provides a one-to-one
correspondence between the members of lattice of cheapest stable
matchings and the non-trivial members of ring-set ${\cal R}_1$.  \eCo

\medskip A polyhedral description of the convex hull of stable
matching was given by Rothblum \cite{Rothblum92}.  This is quite
simple and uses only a small number of inequalities.  In this light,
it is perhaps surprising that the literature, to our best knowledge,
does not know about a result which exhibit this polytope as a member
of integral polyhedra defined by circulations, tensions, submodular
flows, L$_2$-/M$_2$-convex sets.  The following corollary shows that
such an embedding does exist.

\Corollary \label{politop} The polytope of stable matchings of a
bipartite preference system can be obtained as the projection of a
feasible tension polyhedron.  \eCo

\Proof Consider the digraph $D$ associated with $G$ along with the
ring-set ${\cal R}_D$.  Recall that the non-trivial members of ${\cal
R}_D$ are exactly those $s\sp *\ol {t\sp *}$-sets, which are not left
by any dummy arc and left by exactly $n$ stable arcs of $D$.
Furthermore, such leaving arc-sets correspond to the stable matchings
of $G$.  Consider the following polyhedron of feasible potentials.
Let $\Pi := \{\pi :  \pi (s\sp *)=0, \ \pi (t\sp *)=1, \ \Delta _\pi
(e)\leq 1$ \ for every stable arc and $\Delta _\pi (e)\leq 0$ for
every dummy arc$\}$.  Let $\Delta _\Pi$ denote the set of
potential-drops defined by the members of $\Pi$.  Then the vector $\ul
1- \chi (X)$ is in $\Pi$ for every member $X$ of ${\cal R}_D$, and
conversely, for every integer-valued (and hence $(0,1)$-valued)
element of $\Pi$, the set of $0$-valued components is a member of
${\cal R}_D$.

Moreover, by projecting $\Delta _\Pi$ on the set of stable arcs of $D$,
we obtain a polyhedron which is integral (and hence its vertices are
actually $(0,1)$-valued) and its integral elements correspond to the
edge-sets leaving some members of ${\cal R}_D$, which members just
correspond to the stable matchings of $G$.  \FBOX

\medskip

It is well-known that the polytope of perfect matchings of a bipartite graph can be obtained as the projection of a feasible circulation polyhedron. In this light, Corollary \ref{politop} indicates a perhaps surprising difference of the worlds of bipartite stable matchings and perfect matchings.

\Remark \label{TDI1} {\em It is not difficult to read out a min-max
formula from Corollary \ref{politop} for the minimum cost of a stable
matching.  By relying on the weighted version of the theorem of
Dilworth, we develop in Section \ref{poset} (Corollary \ref{wcovering})
a linear system of stable matchings which is TDI (along with an
explicit min-max formula for the maximum weight of a stable matching).
It is important to emphasize that Rothblum \cite{Rothblum92} provided
an linear description of the polytope of stable matchings, which uses
only $O(\vert E\vert )$ linear inequalities.  In addition, Kir\'aly and
Pap \cite{Kiraly+Pap} proved that this linear description of Rothblum
is actually TDI.  It is an interesting challenge to derive the
TDI-ness of the Rothblum system from Corollary \ref{politop}.
$\bullet $ }\eRe

\subsubsection{Multiple cost-function} \label{multip}

Suppose now that we are given not only a single cost-function on the
edge-set of $G$, but $k$:  $c_1,\dots ,c_k$.  We may assume that these
are non-negative.  In the {\bf multiple cost-function stable matching
problem}, we are interested in finding a stable matching which is
cheapest with respect to $c_1$, within this, it is cheapest with
respect to $c_2$, within this, it is cheapest with respect to $c_3$,
and so on.  Similarly to the case $k=1$, this problem can also be
managed with the help of network flows, as follows.

\Corollary\label{multiplecostswithflows}
The multiple cost-function stable matching problem can be solved algorithmically with the help of minimum weight network flows.
\eCo

\Proof
Consider again the digraph $D=(V,A)$ associated with $G$ in Section
\ref{assoc}.  Let $g_i$ denote the capacity function on $A$ assigned
to $c_i$ in the way described in \eref{(gc)}.  Due to Corollary
\ref{code1}, we can compute an $s\sp *\ol{t\sp * }$-set $Z$ which is
not left by any dummy arc and for which $\delta _{g_1}(Z)$ is minimum,
within this $\delta _{g_2}(Z)$ is minimum and so on.  By Theorem
\ref{cgcorres}, the $n$ stable arcs of $D$ leaving $Z$ correspond to a
stable matching of $G$ that minimizes the multiple cost-function
$\{c_1,\dots ,c_k\}$.
\FBOX

In section \ref{fair1}, we will show how this multiple cost stable
matching algorithm can be used to solve a general fair stable matching
problem.

\subsubsection{Forbidden and forced edges} \label{forob}

Let $F \subseteq E_st$ and $N \subseteq E_st$ be two disjoint subsets of edges. We refer to the elements of
$F$ as {\bf forbidden edges}, the elements of $N$ are the {\bf forced edges}.

\Corollary

\eCo

\Proof
Define a cost-function $c_0$ to be 0 on the forced edges, to be $n+1$ on the forbidden edges, and to be $1$ on the other edges.
The expected stable matching $M$ exists if and only if the minimal
$c_0$--cost of a stable matching of $G$ is $n-\vert N\vert $. \FBOX 

In addition, by applying the algorithm outlined above for finding a minimum multiple cost stable matching, we can compute for a given cost-function $c$ a minimum $c$-cost stable matching $M$ for which $N\subseteq M\subseteq E-F$.

By using a different approach, in Corollary \ref{wcovering} we shall
provide a simple characterization for free subsets of edges which
include a stable matching.

\subsection{Packing cheapest stable matchings}

The one-to-one correspondence given in Lemma \ref{corres} between the
stable matchings of a bipartite graph $G$ and stable $s\sp *t\sp
*$-cuts of the associated digraph $D$ can be used not only for finding
a cheapest stable matching but for finding $\ell$ disjoint cheapest
stable matchings, as well.

Here the basic problem aims at finding a maximum number of disjoint
stable matchings.  Ganesh et al.  \cite{GPNP21} developed a linear
time algorithm for this packing problem, however, they did not
consider whether there is here a min-max formula.

\medskip

Here we provide a solution to the problem of finding $\ell$ disjoint
cheapest stable matchings.  When $\ell=1$, this is just the cheapest
stable matching problem.  When $c\equiv 0$, this is just the packing
problem of stable matchings.

\THEOREM \label{mincostpak1} Let $c$ be a non-negative cost-function
on the edge-set of bipartite graph $G=(U,W;E)$ endowed with a
preference system.  The maximum number of disjoint $c$-cheapest stable
matchings is equal to the minimum cardinality of a blocker of
$c$-cheapest stable matchings.  There is a strongly polynomial
two-phase greedy algorithm for computing a largest set of disjoint
$c$-cheapest stable matchings and a minimum cardinality blocker of
$c$-cheapest stable matchings.  \eT

\Proof Consider the correspondence described in Corollary
\ref{mincost} between cheapest stable matchings and the non-trivial
members of the ring-set ${\cal R}_1$ occurring in the corollary.
Based on this, the min-max formula in the theorem is an immediate
consequence of Theorem \ref{pakol} when it is applied to the special
case $h\equiv 1$.  Furthermore, the algorithmic part of the theorem is
a special case of the algorithm described in Section \ref{direkt}.
\FBOX

\Remark {\em At first sight it may seem a bit surprising that the
formal analogue of the min-max formula in Theorem \ref{mincostpak1}
concerning maximum packings of perfect matchings of a bipartite graph
(without a preference system) fails to hold.  That is, it is not true
that in a perfectly matchable bipartite graph the maximum number of
edge-disjoint perfect matchings is equal to the minimum number of
edges blocking all perfect matchings.  To see this, consider the
bipartite graph $G$ consisting of three openly disjoint paths of three
edges connecting nodes $s$ and $t$.  This is an elementary bipartite
graph (that is, every edge belongs to a perfect matching) in which
each perfect matching uses the middle edge of two of the three
$st$-paths.  Therefore $G$ has no two disjoint perfect matchings.  On
the other hand, for every edge $e$ of $G$, there is a perfect matching
avoiding $e$, that is, the perfect matchings cannot be met by a single
edge.

It should be noted that there is a good characterization for the
existence of $\ell$ disjoint perfect matchings of a bipartite graph
(see, for example, Corollary 21.4c in the book of Schrijver
\cite{Schrijverbook}).  $\bullet $ }\eRe

\medskip

By using Theorem \ref{pakol} in its general form, Theorem
\ref{mincostpak1} can be extended as follows.  Let $h$ be a
non-negative integer-valued function on the set of stable edges of
$G$. Recall the definition of h-independence of a family of stable matchings given before Corollary \ref{gindep}. 

\THEOREM \label{mincostpak2} Let $c$ be a cost-function and $h\geq 0$
an integer-valued upper-bound function on the set of stable edges of a
bipartite graph $G$.  The maximum number of $h$-independent
$c$-cheapest stable matchings is equal to the minimum total $h$-value
of a set of stable edges intersecting all $c$-cheapest stable
matchings.  In particular (when $c\equiv 0$), the maximum number of
$h$-independent stable matchings is equal to $$ \hbox{ $\min \{
\widetilde h(L) :  \ L\subseteq E_{\rm st}, \ L$ intersects every
stable matching$\}.$}\ $$

Moreover, a maximum $h$-independent family of $c$-cheapest stable
matchings and a set of edges of minimum total $h$-value intersecting
all cheapest stable matchings can be computed in strongly polynomial
time by a single application of Dijkstra's algorithm.  \FBOX \eT

\section{Fair stable matchings} \label{fair1}

\medskip In the cheapest stable matching problem there was only a
single cost-function and we wanted to minimize the total cost of a
stable matching.  It is a natural requirement to find a stable
matching $M$ that is fair or egalitarian in some sense among the persons
(the nodes of $G$).  For example, one may want to minimize the number
of those persons who get in $M$ their worst stable edge.  To manage
this problem, define a cost-function $c:E_{\rm st}\rightarrow
\{0,1,2\}$, as follows.

\eq c(e) := \begin{cases} 0 & \ \ \hbox{when $e$ is not the worst
stable edge at either of its end-nodes}\ \ \\ 1 & \ \ \hbox{when $e$
is the worst stable edge at exactly one of its end-nodes }\ \ \\ 2 &
\ \ \hbox{when $e$ is the worst stable edge at both of its end-nodes.
}\ \ \ \end{cases} \eeq

For such a $c$, the cost of a stable matching is exactly the number of
those persons who get their worst incident stable edge.  Therefore, a
minimum $c$-cost stable matching is one that minimizes the number of
persons who got their worst stable edge.

This fairness concept, however, is not appropriately sensitive because
it does not take into consideration preferences other than the worst
ones, and there is indeed a rich literature concerning the various
concepts of fairness.  These are discussed in the books of Gusfield
and Irving \cite{Gusfield+Irving} and of Manlove \cite{Manlove-book},
and in a more recent paper of Cooper and Manlove
\cite{Cooper+Manlove2}.  In what follows, we consider
a fairness concept that uses a level-representation of preferences, which is a more refined version of rank-based fairness concepts like rank-maximal stable matching and generous stable matching (see \cite{Cooper+Manlove2} for a discussion of the latter concepts).

Let $L:=\{1,\dots ,\elll\sp * \}$ where $\elll \sp * := 2\vert E_{\rm st}\vert
$. Suppose that at each person (that is, at each node $v$ of $G$) not
only a strict preference list is specified for the edges ending at $v$
but we assign a number $\elll(v,e)\in L$ to the ordered node-edge pairs
$(v,e)$ \ (where $e\in E_{\rm st}$ is an edge incident to $v$) in such a way
that these values at $v$ are distinct and $\elll(v,e) > \elll(v,f)$ if $e$ is
better (at $v$) than $f$. Such a function $\elll$ is called a {\bf level-representation} of the preferences. Hence the values of level-representation $\elll$ are different at any given node $v$, but otherwise they may be equal.  For example, it is allowed for
a stable edge $e=uw$ that $\elll(u,e) = \elll(w,e)$.

For a stable matching $M$, the {\bf $M$-level} $\lambda _M(v)$ of a
node $v\in V$ is defined by $\lambda _M(v):  = \elll(v,e)$ where $e$ is
the element of $M$ incident to $v$.  For a value $\lambda \in L$, we
call a stable edge $e=uw$ {\bf $\lambda $-feasible} if $\elll(u,e) \geq
\lambda $ and $\elll(w,e) \geq \lambda $. We say that stable matching $M$ is
{\bf $\lambda $-feasible} if $M$ consists of $\lambda $-feasible edges,
or equivalently, the $M$-level of each node is at least $\lambda $.

We call a stable matching $M$ {\bf level-fair} (from below) or just {\bf
fair} with respect to level-representation $\elll$ if the number of nodes with
$M$-level $1$ is as small as possible, within this, the number of
nodes with $M$-level 2 is as small as possible, within this, the
number of nodes with $M$-level 3 is as small as possible, and so on.
Our goal is to develop an algorithm for computing a level-fair stable
matching.

\Remark
{\em We remark that level-fair stable matchings, just like rank-maximal stable matchings, can be found in polynomial time using a weighted stable matching algorithm with exponential weights. However, the usage of exponential weights is inconvenient in practical problems. Our aim is to present an algorithm that avoids exponential weights. A discussion of how exponential weights can be efficiently avoided in the rank-maximal stable matching problem can be found in \cite{Cooper+Manlove2}.}
\eRe


We define iteratively a sequence $\lambda _1< \lambda _2< \cdots <
\lambda _k$ of members of $L$ and a sequence $\beta _1, \beta _2,\dots
,\beta _k$ of positive integers for which $\beta _1+\beta _2+\cdots
+\beta _k=2n$.

Let $\lambda _1 \in L$ be the largest value such that there is a
stable matching $M$ for which the $M$-level of every node is at least
$\lambda _1$.  Let ${\cal SM}_1$ denote the set of $\lambda
_1$-feasible stable matchings.  Let $\beta _1$ denote the minimum
number of nodes with $M$-level $\lambda _1$, where the minimum is
taken over all members $M$ of ${\cal SM}_1$.  If $\beta _1=2n$, that
is, if the $M$-level of each node is $\mu _1$ for every $M \in {\cal
SM}_1$, then, by letting $k:=1$, the iterative sequence of definitions
terminates.  (In this case, every member $M$ of ${\cal SM}_1$ consists
of edges $e=uw$ for which $\elll(e,u)=\elll(e,v)=\lambda _1$, and hence ${\cal
SM}_1$ is the wanted set of level-fair stable matchings.)  If $\beta
_1<2n$, we define ${\cal SM}'_1$ to be the set of those members $M$ of
${\cal SM}_1$ for which the number of nodes with $M$-level $\lambda
_1$ is $\beta _1$.

Suppose now that $\lambda _{i-1}$, $\beta _{i-1}$, ${\cal SM}'_{i-1}
\subset {\cal SM}_{i-1}$ have already been defined for a subscript
$i\geq 2$.  Let $\lambda _i \in L$ \ ($\lambda _i> \lambda _{i-1}$) be
the largest value such that there is a stable matching $M\in {\cal
SM}'_{i-1}$ for which the $M$-level of every node of $G$ is either one
of the values $\lambda _1,\dots ,\lambda _{i-1}$ or at least $\lambda
_i$.  Let ${\cal SM}_i$ denote the set of these stable matchings.  Let
$\beta _i$ denote the minimum number of nodes with $M$-level $\lambda
_i$, where the minimum is taken over all members $M$ of ${\cal SM}_i$.


If $\beta _1+\beta _2+\cdots +\beta _i=2n$, then, by letting $k:=i$,
the iterative sequence of definitions terminates.  (In this case, it
holds for every member $M$ of ${\cal SM}_k$ that there are $\beta _i$
nodes of $M$-level $\mu _i$ for $i=1,\dots ,k$ and hence ${\cal SM}_k$
is the wanted set of level-fair stable matchings.)

If $\beta _1+\beta _2+\cdots +\beta _i< 2n$, we define ${\cal SM}'_i$
to be the set of those members $M$ of ${\cal SM}_i$ for which the
number of nodes with $M$-level $\lambda _i$ is $\beta _i$.


\medskip

\noindent {\bf The algorithm for computing a level-fair stable
matching} \ \ Our next goal is to show how the parameters $\mu _i$ and
$\beta _i$ introduced above can be computed for $i=1,\dots ,k$.
Accordingly, the algorithm consists of $k$ stages, each divided into
two halves.

In the first half of Stage 1, we compute $\lambda _1$, as follows.
With subsequent applications of the algorithm outlined in Section
\ref{forob}, we check one-by-one for values $\lambda := \elll\sp *,\elll\sp
*-1,\dots $ whether there is a $\lambda $-feasible stable matching
$M$.  Then $\lambda _1$ is the first $\lambda $ in this sequence for
which a $\lambda $-feasible stable matching exists.

In the second half of Stage 1, we compute a member $M$ of ${\cal
SM}_1$ for which the number of nodes with $M$-level $\lambda _1$ is as
small as possible.  (This minimum number was denoted by $\beta _1$.)
To this end, define a cost-function $c_1$ on stable edges, as follows.

\eq c_1(e) := \begin{cases} 2 & \ \ \hbox{if $\elll(e,v)=\lambda _1$ for
both end-nodes $v$ of $e$ }\ \\ 1 & \ \ \hbox{if $\elll(e,v)=\lambda _1$
for exactly one end-node $v$ of $e$ }\ \\ 0 & \ \ \hbox{otherwise.}\
\end{cases} \eeq

Observe that the $c_1$-cost of a member $M$ of ${\cal SM}_1$ is the
number of nodes with $M$-level $\lambda _1$, and hence $\beta _1= \min
\{\widetilde c_1(M):  M\in {\cal SM}_1\}$.  Therefore, with the help
of the multiple cost-function algorithm outlined in Section
\ref{multip}, $\beta _1$ can be computed.  If $\beta _1=2n$, then
$k=1$ and, as noted above at the definition of $\beta _1$, ${\cal
SM}_1$ is the wanted set of level-fair stable matchings.  If $\beta
_1<2n$, then the second half and hence the whole Stage 1 halts, and
the algorithm turns to subsequent stages, which are analogous to Stage
1.

For describing Stage $i\geq 2$, suppose that the values $\lambda
_1,\dots ,\lambda _{i-1}$ and the values $\beta _1,\dots ,\beta
_{i-i}$ have already been computed, as well as the families ${\cal
SM}'_{i-1} \subseteq {\cal SM}_{i-1}$ of stable matchings.  Similarly
to Stage 1, we can compute in the first half of Stage $i$ the largest
value $\lambda _i\in L$ \ for which there is member $M$ of ${\cal
SM}'_{i-1}$ such that the $M$-level of every node is either one of
$\lambda _1, \lambda _2,\dots ,\lambda _{i-1}$ or at least $\lambda
_i$.

In the second half of Stage $i$, we compute a member $M$ of ${\cal
SM}_i$ for which the number of nodes with $M$-level $\lambda _i$ is as
small as possible.  (This minimum number was denoted by $\beta _i$.)
To this end, define a cost-function $c_i$ on stable edges as follows.

\eq c_i(e) := \begin{cases} 2 & \ \ \hbox{if $\elll(e,v)=\lambda _i$ for
both end-nodes $v$ of $e$ }\ \\ 1 & \ \ \hbox{if $\elll(e,v)=\lambda _i$
for exactly one end-node $v$ of $e$ }\ \\ 0 & \ \ \hbox{otherwise.}\
\end{cases} \eeq

The $c_i$-cost of a member $M$ of ${\cal SM}_i$ is the number of nodes
with $M$-level $\lambda _i$, and hence $\beta _i= \min \{\widetilde
c_i(M):  M\in {\cal SM}_i\}$.  Therefore, with the help of the
multiple cost-function algorithm outlined in Section \ref{multip},
$\beta _i$ can also be computed.  If $\beta _1+\beta _2+\cdots +\beta
_i=2n$, then $k=i$ and ${\cal SM}_k$ is the wanted set of level-fair
stable matchings.  If $\beta _1+\beta _2+\cdots +\beta _i<2n$, then
the second half and hence the whole Stage $i$ halts.

There will be a subscript $i$ for which $\beta _1+\beta _2+\cdots
+\beta _i=2n$ holds in Stage $i$, and this is the moment when the
whole algorithm terminates.  \FBOX

\section{Optimal packing of $st$-cuts of a digraph} \label{cutpack}

Our next goal is to solve the problem of finding $\ell$ disjoint
stable matchings whose union is of minimum cost for a given
cost-function.  This will be discussed only in the next section:  here
we work out the corresponding optimization problem in the digraph
associated with the preference system on $G$.

\subsection{Min-max formula and algorithm}

Let $D=(V,A)$ be a loopless digraph with source-node $s\in V$ and
sink-node $t\in V$, for which we suppose that $\varrho (s)=\delta
(t)=0$.  Let $g:A\rightarrow \ol {\bf Z}_+$ be an integer-valued
capacity-function (allowing value $+\infty $).  Throughout we assume
that $g$ is integer-valued but the presented approach can immediately
be extended to the case when $g$ is rational-valued.

A non-negative function $z:A\rightarrow {\bf R}_+$ is an {\bf
$st$-flow} (or just a flow), if $\varrho _z(v)=\delta _z(v)$ holds for
every node $v\in V-\{s,t\}$.  The flow $z$ is {\bf $g$-feasible}, if
$z\leq g$.  The {\bf amount} of flow $z$ is $\delta _z(s) \ (=\varrho
_z(t)$).  We say that an arc $a\in A$ is {\bf $g$-finite} if $g(a)$ is
finite, while an $st$-cut is {\bf $g$-finite} (or just finite) if each
of its arcs is $g$-finite.  (As before, {\bf $st$-cut} is the set of
arcs leaving an $s\ol t$-set $Z\subset V$.

Let $\ell$ be a positive integer, and suppose that there are $\ell$
arc-disjoint $g$-finite $st$-cuts, which is equivalent to requiring
that every $st$-path contains at least $\ell$ $g$-finite arcs.  (This
problem is nothing but a shortest path problem.)

We call the union of $\ell$ arc-disjoint $st$-cuts an {\bf
$\ell$-cut}.  The {\bf $g$-capacity} (or just the capacity or the
$g$-value) of an $\ell$-cut $L\subseteq A$ is $\widetilde g(L)$, that
is, the sum of $g$-capacities of the $st$-cuts in $L$.  The system of
$s\ol t$-sets $Z_1,\dots ,Z_\ell$ is {\bf arc-independent} if the
$\ell$ $st$-cuts defined by these sets are arc-disjoint.

The major problem of this section is finding and characterizing
$\ell$-cuts with minimum $g$-capacity.  In the special case $\ell=1$,
This is answered by the MFMC theorem.  For a min-max formula
concerning the general case $\ell\geq 1$, consider an integer-valued
$st$-flow $z$, which may not be $g$-feasible.  We call an arc $a$ {\bf
overloaded} if $z(a)>g(a)$.  The {\bf surplus} of arc $a\in A$ is:  $$
\hbox{ $\omega _z(a) := \ (z(a)-g(a))\sp +$, }\ $$ where $x\sp +:=
\max \{x,0\}.$ The {\bf surplus} $\omega (z)$ of a flow $z$ is the sum
of the surpluses of its arcs, that is, $$ \omega (z):  = \sum
[(z(a)-g(a))\sp +:  a\in A].  $$

\THEOREM \label{main1} Suppose that the digraph $D=(V,A)$ admits a
$g$-finite $\ell$-cut (which is the union of $\ell$ arc-disjoint
$g$-finite $st$-cuts), or equivalently, every $st$-path has at least
$\ell$ $g$-finite arcs.  Then:  \medskip

\noindent {\bf (A)}

\eq \hbox{ $ \min \{\widetilde g(L):  L \ (\subseteq A)$ \ an \
$\ell$-cut$\} \ =$ }\ \label{(lmin)} \eeq \eq \hbox{ $ \max \{ \ell
\delta _z(s) - \omega (z) :  \ z$ \ an integer-valued $st$-flow$\}.$
}\ \label{(lmax)} \eeq

\noindent Moreover, both a $\widetilde g$-minimizer $\ell$-cut $L$ in
\eref{(lmin)} and an integral $st$-flow $z$ maximizing \eref{(lmax)}
can be computed in strongly polynomial time with the help of the
min-cost flow algorithm of Ford and Fulkerson.  \medskip

\noindent {\bf (B)} \ An $\ell$-cut $L$ defined by arc-independent
$s\ol t$-sets $Z_1, \dots ,Z_\ell$ is an optimal solution to
\eref{(lmin)} if and only if there exists an integer-valued $st$-flow
$z$ for which the following optimality criteria hold for every arc
$a\in A$.

\eq \begin{cases} {\bf {\rm {\bf (O1)}} } \ \ \ \ \ z(a) > g(a) \ \ \
\ \ \Rightarrow \ \ \ \ \ a\in L \ \ \ \ \\ {\rm {\bf (O2)}} \ \ \ \
\ z(a) < g(a) \ \ \ \ \ \Rightarrow \ \ \ \ \ a\in A-L \\ {\rm {\bf
(O3)}} \ \ \ \ \ z(a) > 0 \ \ \ \ \ \ \ \ \ \Rightarrow \ \ \ \ \ a \
\ \hbox{does not enter any} \ Z_i \hbox{}.  \end{cases}
\label{(optkrit)} \eeq

\noindent {\bf (C)} \ The $\ell$-cut minimizing \eref{(lmin)} may be
chosen in such a way that its defining $\ell$ arc-independent $s\ol
t$-sets form a chain.  \eT

\Proof We start with the proof of Part \ {\bf (C)}.

For a system $\cal F$ of $s\ol t$-sets, let $\widetilde \delta
_g({\cal F}) := \sum [\delta _g(X):  X \in {\cal F}]$.  Consider an
$\ell$-cut minimizing \eref{(lmin)} for which the square-sum $\sum [
\vert X\vert \sp 2:  X \in {\cal F}]$ of its defining arc-independent
set-system ${\cal F}$ is minimum.  We claim that $\cal F$ is a chain.

Suppose indirectly that $\cal F$ has two members $X$ and $Y$ for which
$X-Y$ and $Y-X$ are non-empty.  Then the modified set-system ${\cal
F}':= {\cal F} - \{X,Y\} \cup \{X\cap Y, X\cup Y \}$ is also
arc-independent, for which $\delta _g(X) + \delta _g(Y) \geq \delta
_g(X\cap Y) + \delta _g(X\cup Y)$ implies that $\widetilde \delta
_g({\cal F}) \geq \widetilde \delta _g({\cal F}') \geq \widetilde
\delta _g({\cal F})$, and hence $ \widetilde \delta _g({\cal F}') =
\widetilde \delta _g({\cal F})$ follows that is, the $\ell$-cut
defined by ${\cal F}'$ is also a minimizer of \eref{(lmin)},
contradicting the assumption that the square-sum of $\cal F$ is
minimum.

We note that this simple direct proof helps understanding the
structure of smallest $\ell$-cuts but actually we do not really need
it since the algorithmic proof below for the min-max formula provides
automatically such a chain.

\medskip

In order to prove {\bf (A)}, we consider first the easier direction
$\max \leq \min$.  \ To this end, let $L\subseteq A$ be an $\ell$-cut
\ which is the set of arcs leaving one of the members $Z_1,\dots
,Z_\ell$ of an arc-independent system of $s\ol t$-sets.  Let $A_i$
denote the set of arcs leaving $Z_i$, that is, $L$ is the union of the
$\ell$ disjoint sets $A_i$.  Furthermore let $z$ be an $st$-flow, and
let $A_> := \{a\in A:  z(a)> g(a)\}$ and $A_\leq := \{a\in A:
z(a)\leq g(a)\}$.  Then

\eq \begin{cases} \widetilde g(A_i) = \widetilde g(A_i\cap A_\leq ) +
\widetilde g(A_i\cap A_>) \\ \geq \widetilde z(A_i\cap A_\leq ) +
\widetilde g(A_i\cap A_>) \\ = \widetilde z(A_i\cap A_\leq ) +
\widetilde z(A_i\cap A_>) - [ \widetilde z(A_i\cap A_>) - \widetilde
g(A_i\cap A_>)] \\ = \widetilde z(A_i) - [ \widetilde z(A_i\cap A_>)
- \widetilde g(A_i\cap A_>)], \end{cases} \label{(becsles1)} \eeq and
here equality holds if and only if $\widetilde z(A_i\cap A_\leq ) =
\widetilde g(A_i\cap A_\leq )$.  From this estimation, we obtain the
following.

\eq \begin{cases} \widetilde g(L) = \sum _{i=1}\sp \ell \widetilde
g(A_i) \\ \geq

\sum _{i=1}\sp \ell \widetilde z(A_i) - [ \widetilde z(A_i\cap A_>) -
\widetilde g(A_i\cap A_>)] \\ \geq

\sum _{i=1}\sp \ell \widetilde z(A_i) -\omega (z) = \sum _{i=1}\sp
\ell \delta _z(Z_i) -\omega (z) =

\ell \delta _z(s) - \sum _{i=1}\sp \ell \varrho _z(Z_i) -\omega (z)
\\ \geq

\ell \delta _z(s) - \omega (z), \cr

\end{cases} \label{(estim)} \eeq from which the inequality $\max \leq
\min$ follows.

\medskip

To prove the non-trivial inequality $\max \geq \min$ for {\bf (A)} and
statement \ {\bf (B)}, \ we derive first the following.

\Claim \label{ekviv} In the estimation \eref{(estim)}, equality holds
throughout if and only if each of the three optimality criteria in
\eref{(optkrit)} is met.  \eCl

\Proof The first inequality in \eref{(estim)} holds with equality if
and only if $\widetilde z(A_i\cap A_\leq ) = \widetilde g(A_i\cap
A_\leq )$ holds for each $i=1,\dots ,\ell$, that is, $z(a)\geq g(a)$
holds for each arc $a$ leaving some $Z_i$, and this is exactly
Optimality criterion ({\bf O2}).  The second inequality holds with
equality if and only if every overloaded arc is in $L$, and this is
exactly Optimality criterion ({\bf O1}).  Finally, the third
inequality holds with equality if and only if $z(a)=0$ holds for every
arc $a$ entering some $Z_i$, and this is exactly Optimality criterion
({\bf O3}).  \FBOX

\medskip

Next, we construct an integer-valued $st$-flow $z\sp *$ along with an
$\ell$-cut $L\sp *$ for which the defining set-system ${\cal F}\sp *$
consisting of arc-independent $s\ol t$-sets $Z_1,\dots ,Z_\ell$ is a
chain, and each inequality in the estimation \eref{(estim)} is met by
equality, and hence, by Claim \ref{ekviv}, the optimality criteria in
\eref{(optkrit)} hold.

For each $g$-finite arc $e\in A$, add a parallel copy $e'$.  Let $A'$
denote the set of these new arcs and let $A_1:=A\cup A'$.  Define the
capacity-function $g_1 :  A_1\rightarrow {\bf \ol Z}_+$ and the
cost-function $c_1 :A_1 \rightarrow \{0,1\}$ as follows.

\eq g_1(e):= \begin{cases} g(e) & \ \ \hbox{if}\ \ \ e\in A \cr
+\infty & \ \ \hbox{if}\ \ \ e \in A', \end{cases} \eeq

\eq c_1(e):= \begin{cases} 1 & \ \ \hbox{if}\ \ \ e\in A' \\ 0 & \ \
\hbox{if}\ \ \ e\in A. \end{cases} \eeq

Consider the classic Ford-Fulkerson algorithm \cite{Ford-Fulkerson}
for computing a $c_1$-cheapest $g_1$-feasible integral $st$-flow in
digraph $D_1:=(V, A_1)$.  (See also the version of the algorithm
outlined in pages 128-129 of the book \cite{Frank-book}.)  Recall that
the cost-function $c_1$ is $(0,1)$-valued, and the capacity-function
$g_1$ is integer-valued.

At a given stage of the run of the algorithm, we have at hand a
current integer-valued potential $\pi \geq 0$ defined on $V$, for
which $\pi (s)=0$, along with a $g_1$-feasible $st$-flow $z:
A_1\rightarrow {\bf Z} _+ $ meeting the following optimality criteria.

\eq \begin{cases} {\bf {\rm {\bf (F1)}} } \ \ \ \ \ c_1(e) > \Delta
_\pi (e) \ \ \ \ \ \Rightarrow \ \ \ \ \ z(e) = 0 \ \ \ \ \\ {\rm
{\bf (F2)}} \ \ \ \ \ c_1(e) < \Delta _\pi (e) \ \ \ \ \ \Rightarrow \
\ \ \ \ z(e)=g_1(e), \end{cases} \label{(foptkrit)} \eeq where $\Delta
_\pi (e):=\pi (v)-\pi (u)$ for arc $e=uv\in A_1$.

At the beginning, $\pi \equiv 0$ and $z\equiv 0$.  In the procedure,
two kinds of phases alternately follow each other:  $\pi $-augmenting
and $z$-augmenting phases.  In a $\pi $-augmenting phase, we increase
the $\pi $-value of certain nodes by 1, without changing the current
flow $z$, in such a way that the optimality criteria continue to hold,
and the value of $\pi (t)$ is increased by 1 at each $\pi $-augmenting
step.

In a flow-augmenting phase the current potential $\pi $ remains
unchanged.  This $\pi $ and the current flow $z$ define an auxiliary
digraph in a standard way.  With the help of a shortest $st$-path in
the auxiliary digraph, we increase the flow-amount as much as
possible.  As Edmonds+Karp and Dinits proved, after at most $O(\vert
V\vert \vert A\vert )$ such flow augmentations the flow-augmenting
phase terminates and we turn to the next potential augmenting phase.

The whole algorithm terminates when the current value of $\pi (t)$
reaches $\ell$.  Let $\pi _1$ denote this final potential, while the
current flow at this moment is denoted by $z_1$.  Recall that $\pi _1$
and $z_1$ meet the Optimality criteria \eref{(foptkrit)}.

Let $K$ denote the flow-amount of $z_1$.  Then $z_1$ is a
$c_1$-cheapest flow in digraph $D_1$ among the $g_1$-feasible flows of
amount $K$.  Note that $\pi _1$ and $z_1$ have been obtained after
$\ell$ potential augmentations and $\ell +1 $ maximum flow
computations, that is, in strongly polynomial time.

If $g(a)$ is finite and $z_1(a') \geq 1$, for an arc $a\in A$, then
$z_1(a)=g(a)$, since if we had $z_1(a)\leq g(a)-1$, then decreasing
$z_1(a')$ by 1 and increasing $z_1(a)$ by 1, we would obtain another
flow of amount $K$ whose $c_1$-cost would be smaller (by $1$) than
that of $z_1$.

Define the function $z\sp *$ on arc-set $A$, as follows.

\eq z\sp *(a) := \begin{cases} z_1(a) & \ \ \hbox{if}\ \ \
g(a)=+\infty \\ z_1(a) + z_1(a') & \ \ \hbox{if}\ \ \ g(a)< +\infty .
\end{cases} \eeq

\noindent Then $z\sp *$ is a flow in $D$ with flow-amount $K$.

Let $Z_i :=\{v\in V:  \pi _1(v)\leq i-1\}$ \ ($i=1,\dots ,\ell$).
Then $s\in Z_1\subseteq \cdots \subseteq Z_\ell\subseteq V-t$ form a
chain ${\cal F}\sp *$ of sets.  Let $L\sp *$ denote the set of arcs of
$D$ leaving the members of chain ${\cal F}\sp *$.

\Claim The members of ${\cal F}\sp *$ are arc-independent in $D$ (that
is, each arc $a=uv\in A$ leaves at most one member), in particular,
the members of ${\cal F}\sp *$ are distinct.  \eCl

\Proof The arc-independence of ${\cal F}\sp *$ is equivalent to
requiring that $\Delta _{\pi _1}(a) = \pi _1(v)-\pi _1(u)\leq 1$ for
every arc $a=uv \in A$.  But this holds indeed since if $g(a)=+\infty
$, then $z_1(a) < +\infty =g(a) = g_1(a)$, and by ({\bf F2}), we have
$\pi _1(v)\leq \pi _1(u)$, that is, such an arc $a$ cannot leave any
$Z_i$.  If in turn $g(a)<+\infty $, then $a'\in A'$ and hence $z_1(a')
< +\infty =g_1(a')$.  By ({\bf F2}), we have $1= c_1(a') \geq \Delta
_{\pi _1}(a') = \pi _1(v)-\pi _1(u)$.  \FBOX \medskip

\Claim The $\ell$-cut $L\sp *$ and the flow $z\sp *$ meet the
optimality criteria \eref{(optkrit)} of the theorem.  \eCl

\Proof To prove ({\bf O1}), suppose that $z\sp *(a) > g(a)$ for some
arc $a= uv\in A$.  Then $z_1(a')\geq 1$, and hence, by relying on
({\bf F1}), we have $1=c_1(a') \leq \Delta _{\pi _1}(a') = \pi _1(v)
-\pi _1(u)$.  Therefore arc $a$ does indeed leave some $Z_i$, that is,
$a$ is in $L\sp *$, and thus ({\bf O1}) holds.

To prove ({\bf O2}), assume that $z\sp *(a) < g(a)$ holds for some arc
$a= uv\in A$.  Then ({\bf F2}) implies that $0=c_1(a) \geq \Delta
_{\pi _1}(a) = \pi _1(v) -\pi _1(u)$, that is, arc $a$ does not leave
any $Z_i$, showing that $a$ is in $A-L\sp *$, and hence ({\bf O2})
holds.

To prove ({\bf O3}), assume that $z\sp *(a) \ (=z_1(a)) > 0$. holds
for some arc $a\in uvA$.  Then ({\bf F1}) implies $0=c_1(a) \leq
\Delta _{\pi _1}(a) = \pi _1(v) -\pi _1(u)$, that is, $\pi _1(v)\geq
\pi _1(u)$, and hence $a$ cannot enter any set $Z_i$, implying that
({\bf O3}) holds.  \FBOX \medskip

Summing up, we proved that the Ford+Fulkerson algorithm for computing
a cheapest feasible flow constructs an $\ell$-cut $L\sp *$ and an
integral $st$-flow $z\sp *$ which meet the optimality criteria
\eref{(optkrit)}, proving in this way the non-trivial inequality $\max
\geq \min$.

In addition, when the strongly polynomial maximum flow algorithm of
Edmonds+Karp or Dinits is used as a subroutine, the Ford+Fulkerson
algorithm is strongly polynomial since the cost-function in question
is $(0,1)$-valued.  \BB

\medskip

\subsection{Packing $st$-cuts defined by a ring-set} \label{st-ring}

Theorem \ref{main1} has a self-refining nature in the sense that it
easily implies the following extension.  Let $\cal R$ be a ring-set
containing $\emptyset $ and $V$ whose non-trivial members are $s\ol
t$-sets.  We assume that $\cal R$ is described by its {\bf
code-digraph} a $D_{\cal R}=(V, A_{\cal R})$ where $uv$ is an arc of
the code-digraph if $u\in V-t$ and $uv$ does not leave any member of
$\cal R$, or equivalently, node $v$ is in the (unique) minimal member
of $\cal R$ containing $u$.  We call a $g$-finite $st$-cut of $D$ {\bf
${\cal R}$-compatible} if its out-shore is a member of $\cal R$, while
a $g$-finite arc-set $L\subseteq A$ is an {\bf ${\cal R}$-compatible
$\ell$-cut} if it is the disjoint union of $\ell$ ${\cal
R}$-compatible $st$-cuts.  We are interested in finding an ${\cal
R}$-compatible $\ell$-cut $L$ for which $\widetilde g(L)$ is minimum.

To manage this problem, extend function $g$ (originally defined on the
arc-set of $D$) to the code-arcs of $\cal R$ by defining it $+\infty $
on each code-arc.  Let $D\sp +=(V,A\sp +)$ denote the digraph obtained
from $D$ by adding each code-arc.  Then every $s\ol t$-set which is
not in $\cal R$ admits a leaving arc with capacity $+\infty $.

By applying Theorem \ref{main1} to $D\sp +$, we obtain the following
min-max formula.

\Corollary \label{main1b} Let $D=(V,A)$ be digraph endowed with a
non-negative, integer-valued function $g$ on $A$.  Let $\cal R$ be a
ring-set (given by its code-digraph) containing $\emptyset $ and $V$
whose non-trivial members are $s\ol t$-sets.  Let $D\sp +$ denote the
digraph obtained from $D$ by adding the arcs of the code-digraph of
$\cal R$.  We assume that $D$ has a $g$-finite ${\cal R}$-compatible
$\ell$-cut.  Then

\eq \hbox{ $ \min \{\widetilde g(L):  L \ (\subseteq A)$ \ an ${\cal
R}$-compatible $\ell$-cut$\} \ =$ }\ \label{(lminb)} \eeq \eq \hbox{ $
\max \{ \ell \delta _z(s) - \omega (z) :  \ z$ \ an integer-valued
$st$-flow in $D\sp +\}.$ }\ \label{(lmaxb)} \eeq

\noindent Moreover, both a $\widetilde g$-minimizer $\ell$-cut $L$ in
\eref{(lminb)} and an integral $st$-flow $z$ maximizing \eref{(lmaxb)}
can be computed in strongly polynomial time with the help of the
min-cost flow algorithm of Ford and Fulkerson.  \FBOX \eCo

\section{Packing and covering problems of stable matchings}

\subsection{Disjoint stable matchings with minimum total cost}
\label{pack2.1}

Theorem \ref{mincostpak1} provided an answer to the problem of finding
$\ell$ disjoint minimum c-cost stable matchings.  As a natural
generalization, one may be interested in finding $\ell$ disjoint
stable matchings for which the $c$-cost of their union is minimum with
respect to a rational cost-function $c$.  We may assume that $c$ is
non-negative, and that $c$ is actually integer-valued.

\THEOREM \label{mincostpakx} Assume that the bipartite graph $G$ has
$\ell$ disjoint stable matchings.  With the help of a min-cost flow
algorithm (as described in Section \ref{olcsost}), it is possible to
compute $\ell$ disjoint stable matchings whose union has minimum
$c$-cost.  \eT

\Proof Consider the digraph $D$ (with source-node $s\sp *$ and
sink-node $t\sp *$) associated with the preference system on $G$ in
Section \ref{assoc}, as well as the function $g_c$ on $A$ associated
with the cost-function $c$ on the set of stable edges of $G$.  Recall
the definition of ring-set ${\cal R}_D$ and Corollary \ref{code1}
which established a one-to one correspondence between the stable
matchings of $G$ and the ${\cal R}_D$-compatible $s\sp *t\sp *$-cuts
of $D$.

In this correspondence, the $c$-cost of a stable matching was equal to
the $g_c$-capacity of the corresponding ${\cal R}_D$-compatible $s\sp
*t\sp *$-cut of $D$.  Based on this, Section \ref{olcsost} described
how a minimum $c$-cost stable matching can be computed by an MFMC
algorithm that computes a minimum $g_c$-capacity ${\cal
R}_D$-compatible $s\sp *t\sp *$-cut of $D$.

Exactly the same correspondence shows that the problem of finding
$\ell$ disjoint stable matchings whose union is of minimum $c$-cost
can be solved by computing $\ell$ disjoint ${\cal R}_D$-compatible
$s\sp *t\sp *$-cuts of $D$ for which the $g_c$-value of their union is
minimum.  But such an algorithm was described in Section
\ref{st-ring}.  \FBOX

\medskip

\subsection{Maximum weight union of stable matchings} \label{pack2.2}

Suppose now that $w$ is a non-negative integer-valued function on
$E_{\rm st}$ and we are interested in finding $\ell$ not necessarily
disjoint stable matchings of a preference system on $G=(U,W;E)$ whose
union is of maximum $w$-weight.  Since each stable matching has the
same cardinality, the version of this problem when the $\ell$ stable
matchings are required to be disjoint, is equivalent to the cheapest
packing problem discussed in Section \ref{pack2.1}.  Also, we may
actually assume that the weight-function $w$ is actually strictly
positive.

To manage the general case when disjointness is not expected, we
introduce the operation of adding a parallel edge to the preference
system.  For a stable edge $e$ of $G$, let $e'$ be a new edge which is
parallel to $e$.  We define $e$ to be girl-better and boy-worse than
$e'$, while the preference relations of $e'$ to other edges in $E-e$
is the same as the ones of $e$.  (That is, if $e$ is, for example,
girl-better than $f$, then $e'$ is also girl-better than $f$).

Let $G'=(U,W;E')$ denote the bipartite graph arising from $G$ by
adding $\ell-1$ edges parallel to $e$ for each stable edge $e$ of $G$.
Define a weight-function $w'$ on $E'$ by letting $w'(e):=w(e)$ for
each original stable edge and $w'(e'):=0$ for a new edge $e'$.

\Lemma \label{non-d-union} The maximum $w$-weight of the union of
$\ell$ (not necessarily disjoint) stable matchings of $G$ is equal to
the maximum $w'$-weight of the union of $\ell$ disjoint stable
matchings of $G'$.  \eL

\Proof Consider first $\ell$ disjoint stable matchings $M'_1,\dots
,M'_{\ell}$ of $G'$ whose union $L'$ has maximum $w'$-weight.  Let
$M_i$ denote the stable matching of $G$ corresponding to $M'_i$ and
let $L$ denote the union of these $M_i$'s.  Since $w$ is strictly
positive, it follows that if a parallel copy $e'$ of a stable edge $e$
is in $M'_i$, then $e$ is in $M'_i$.  Therefore $L$ is the union of
$\ell$ stable matchings of $G$ for which $\widetilde w(L) = \widetilde
w'(L')$.

Second, let $L$ be the union of stable matchings $M_1,M_2,\dots
,M_\ell$ of $G$.  These determine $\ell$ disjoint stable matchings
$M'_1,M'_2,\dots ,M'_\ell$ of $G'$ with union $L'$ for which
$\widetilde w'(L')=\widetilde w(L)$, from which the claim follows
\FBOX.

\medskip By Lemma \ref{non-d-union}, the algorithmic approach
formulated in Theorem \ref{mincostpakx} gives rise to the following.

\Corollary\label{weightedstableunionmax}
Let $w$ be a non-negative weight-function on the set of stable edges of
a bipartite graph endowed with a preference system. With the help of a min-cost MFMC
algorithm, it is possible to compute $\ell$ (not-necessarily disjoint) stable matchings whose union is of maximum $w$-weight.\FBOX
\eCo

\section{Posets and stable matchings} \label{poset}

\medskip

In Section \ref{assoc}, we described a ring-set ${\cal R}_D$ on a
digraph $D$ which encoded the set of stable matchings.  Here we show
that there is a poset on the set $E_{\rm st}$ of stable edges of $G$
which directly captures the main structural properties of stable
matchings.  With this link, we can apply theorems (and algorithms)
concerning posets, such as the ones of Dilworth, Mirsky, and
Greene+Kleitman.

Throughout this section $P=(S,\prec )$ is a poset.  Recall that Dilworth's theorem \cite{Dilworth50}
stated that the maximum cardinality $\alpha :=\alpha (P)$ of an
antichain of $P$ is equal to the minimum number of chains covering
$S$.  A maximum cardinality antichain is called a {\bf
Dilworth-antichain} or, in short, a {\bf D-antichain}.

Let ${\cal C}:= \{C_1,\dots ,C_\alpha $\} be a smallest partition of
$S$ into chains ensured by Dilworth's theorem.  Clearly, a D-antichain
contains exactly one element from each $C_i$.  For two D-antichains
$A_1$ and $A_2$, their {\bf join} $A_1\vee A_2$ ({\bf meet} $A_1\wedge
A_2$) consists of the largest (smallest) elements of $A_1\cup A_2$.
It is well-known that these are D-antichains for which an element of
$C_i\cap A_1\cap A_2$ belongs to both the join and the meet, while if
$C_i$ contains two distinct elements of $A_1\cup A_2$, then the larger
one is in $A_1\vee A_2$ and the smaller one is in $A_1\wedge A_2$.
This implies that there is a unique lowest and a unique highest
D-antichain of $P$.

Mirsky's theorem (sometimes called the polar-Dilworth theorem) states
the maximum cardinality $\gamma _1$ of a chain is equal to the minimum
number of antichains covering $S$.  The theorem of Greene+Kleitman
\cite{Greene-Kleitman} is a min-max formula for the maximum
cardinality of the union of $\ell$ antichains (see Theorem \ref{GK}
below).  In the special case $\ell=1$, this gives back Dilworth, while
in the special case $\ell=\gamma _1$, this gives back Mirsky.

It should also be emphasized that Mirsky's theorem has a simple
algorithmic proof (based on a two-phase greedy algorithm:  see below).
Dilworth's theorem also has an elegant algorithmic proof (due to
Fulkerson \cite{Fulkerson56}) which is based on a reduction to K{\H
o}nig's min-max theorem on maximum matchings.  We note that the
D-antichain obtained by this algorithm is the unique lowest (or
highest) D-antichain.  For the Greene+Kleitman theorem, Frank
\cite{FrankJ5} provided an algorithmic proof based on the min-cost flow
algorithm of Ford and Fulkerson \cite{Ford-Fulkerson}.

\subsection{D-antichain-extendible posets}

We call a poset {\bf D-antichain-extendible} if every maximal
antichain is a D-antichain, or equivalently, every antichain can be
extended to a D-antichain.

\Lemma \label{D-ext} An antichain $A$ of a poset $P=(S,\preceq )$ can
be extended to a $D$-antichain if and only if every subset of $A$ with
at most two elements can be extended to a $D$-antichain of $P$. A poset is
D-antichain-extendible if and only if every antichain with at most two
elements can be extended to a D-antichain.  \eL

\Proof The second half follows immediately from the first.  The
necessity of the condition of the first part is immediate.  To prove sufficiency, we may assume that $\vert A\vert \geq 3$ and every
proper subset of $A$ can be extended to a D-antichain.  Let $\alpha $
denote the cardinality of a D-antichain and $\{C_1,\dots ,C_\alpha \}$
a partition of $S$ into chains.  For $i=1,2,3$, let $a_i$ denote the
single element of $A\cap C_i$.

By the assumption, $A-a_i$ can be extended to a D-antichain $A'_i$ for
each $i=1,2,3$.  If $a_i\in A'_i$ for some $i=1,2,3$, then we are
done, so suppose that this is not the case.  Let $b_i$ denote the
single element of $C_i\cap A'_i$.

As $b_i$ is comparable with $a_i$ for each $i=1,2,3$, there are two
among these subscripts, say $i=1,2$, for which the order relation
between $a_1$ and $b_1$ is the same as the one between $a_2$ and
$b_2$.  So we may assume that $a_1\prec b_1$ and $a_2\prec b_2$.  But
then the meet $A'_1\wedge A'_2$ is a D-antichain including the whole
$A$.  \FBOX

\medskip

Consider again the bipartite graph $G=(U,W;E)$ endowed with a
preference system on its edge-set, and define the {\bf $G$-induced
poset} $P_G:=(E_{\rm st},\preceq )$, as follows.  For two distinct
(though not necessarily incident) stable edges $e$ and $f$ of $G$, we
say that $e$ is larger than $f$ in $P_G$, in notation, $e \succ f$ if
$f$ is (strictly) girl-better than that edge in $M_e$ which is
incident to $f$ in $W$, where $M_e$ is the girl-best stable matching
containing $e$.  (In particular, if a stable edge $f$ \ ($\not =e$) is
girl-better than $e$, then $e\succ f$.)  Observe that if $e \succ f$,
then, for any stable matching $M$ containing $e$, $f$ is (strictly)
girl-better than that edge in $M$ which is incident to $f$ in $W$.

\Lemma \label{ext} The relation $\preceq $ on the elements of $E_{\rm
st}$ is transitive and antisymmetric, that is, $P_G$ is a poset.
Moreover, $P_G$ is D-antichain-extendible.  \eL

\Proof Lemma \ref{corres} implies the following.

\Claim \label{posekvi} For distinct stable edges $e$ and $f$ of $G$,
$e \succ f$ holds if and only if the head of the stable arc $a_f$ of
$D$ (associated with $f$) belongs to $L(M_e)$.  \FBOX \eCl

The claim immediately implies the first part of the lemma.

\Claim \label{uncomp} Two distinct stable edges $e$ and $f$ are
uncomparable in $P_G$ if and only if there is a stable matching
containing both.  \eCl

\Proof Suppose first that $e$ and $f$ are comparable, say $ e \succ
f$.  If, indirectly, there were a stable matching $N$ containing both
$e$ and $f$, then $M_e\wedge N$ (consisting of the girl-best elements
of $M_e\cup N$) would be a stable matching containing $e$ which is
girl-better than $M_e$ contradicting the definition of $M_e$.

Suppose now that no stable matching contains both $e$ and $f$, in
particular $e\not \in M_f$ and $f\not \in M_e$, and indirectly $e$ and
$f$ are not comparable in $P_G$.  Then the head of arc $a_f$ is not in
$L(M_e)$, from which the tail of $a_f$ is not in $L(M_e)$ either, and
analogously, neither the head nor the tail of $a_e$ is in $L(M_f)$.
But then $M_e \vee M_f$ is a stable matching containing both $e$ and
$f$.  \FBOX

\medskip By applying Lemma \ref{D-ext} to poset $P_G$, we obtain that
$P_G$ is indeed D-antichain-extendible.  \BB

\medskip

\Remark {\em We note that a fundamental tool of the book of Gusfield
and Irving \cite{Gusfield+Irving} to manage structural and
optimization problems of stable matchings is a certain poset
$\Pi({M})$ associated with a bipartite preference system.  For
example, their Theorem 3.4.2 characterizes matchings of $G$ which can
be extended to a stable matchings.  It should, however, be emphasized
that their poset is different from the present $P_G$ since the
ground-set of $\Pi({M})$ is the set of rotations while the ground-set
of $P_G$ is $E_{st}$.  $\bullet $ }\eRe

\Remark {\em It should be emphasized that the above concept of induced
poset can be extended to stable $b$-matchings and even to matroid
kernels, implying that the solutions of the optimization problems
discussed in the rest of this section can be extended to those
concerning matroid kernels.  These will be worked out in a forthcoming
paper \cite{FFK2}.  $\bullet $ }\eRe

\medskip

We call a subset $K$ of stable edges {\bf anti-stable} if no two
elements of $K$ belong to the same stable matching.  (In the
literature, an anti-blocker of a set-system $\cal F$ is a subset of
the ground-set that intersects each member of $\cal F$ in at most one
element.  Therefore, $K\subset E_{\rm st}$ is anti-stable precisely if
it is an anti-blocker of the set of stable matchings).  Claim
\ref{uncomp} shows that a set $K\subseteq E_{\rm st}$ is anti-stable
if and only if $K$ is a chain of poset $P_G$, which is equivalent to
requiring that the stable matchings $M_e$ ($e\in K)$ form a chain in
the distributive lattice of stable matchings.

\subsection{Dilworth and Mirsky}

In this section we discuss algorithmic approaches to the weighted
versions of theorems of Mirsky and Dilworth.

\subsubsection{Weighted Mirsky}

The theorem of Mirsky immediately implies its weighted version (see,
Theorem 14.3 in the book of Schrijver \cite{Schrijverbook} or Theorem
2.4.30 in \cite{Frank-book}), which is as follows.

\THEOREM [weighted Mirsky] \label{fMirsky} Given a non-negative
integer-valued function $f$ on the ground-set $S$ of a poset $P$, the
minimum number of antichains covering $f$ is equal to the maximum
$f$-value of a chain.  The optimal $f$-covering family of antichains
can be chosen in such a way that the number of distinct antichains is
at most $\vert S\vert $. There is a two-phase greedy algorithm (see,
\cite{Frank-book} Page 102) which computes such a minimum family of
antichains in the first phase and a chain with maximum total $f$-value
in the second.  When $P$ is D-antichain-extendible, the $f$-covering
antichains may be chosen D-antichains.  \FBOX \eT

For completeness, we outline the two-phase greedy algorithm cited in
the theorem.  The first phase computes an $f$-covering family $\cal A$
of antichains that contains at most $|S|$ distinct antichains.  Let
$f_0:=f$.  In Step $i$ ($i=1,2,\dots $), define $A_i$ to be the set of
minimal $f_{i-1}$-positive elements of $P$.  Let $\mu _i:=
\min\{f_{i-1}(s):  s\in A_i\}$, and define $f_i$ by

\eq f_{i}(s):= \begin{cases} f_{i-1}(s)-\mu _i & \ \ \hbox{if}\ \ \
s\in A_i \\ f_{i-1}(s) & \ \ \hbox{if}\ \ \ s\in S-A_i.  \end{cases}
\eeq

Phase 1 terminates when the current weight-function becomes
identically zero.  Obviously, the sets $A_1,\dots ,A_k$ defined in
Phase 1 are antichains.  Let the family $\cal A$ consist of $\mu _i$
members of $A_i$ ($i=1,\dots ,k$).  Then $\cal A$ consists of $\sum \mu
_i$ antichains and each element $s$ of $P$ belongs to exactly $f(s)$
members of $\cal A$.

In Phase 2, we proceed backward on the antichains $A_k,A_{k-1},\dots
,A_1$.  Select first an arbitrary element $p_1$ of $A_k$.  Let $i$ be
the largest subscript for which $A_i$ does not contain $p_1$ (if there
is any).  Since $p_1$ is not in $A_i$ but it is in $A_{i+1}$, there is
an element $p_2$ of $A_i$ which is smaller than $p_1$.  Continuing in
this way, we are building a chain $C=\{p_1,\dots ,p_t\}$ until the
construction cannot be continued since $p_t$ is in $A_1$.  It follows
that $\widetilde f(C)= \sum [\mu _i:  i=1,\dots ,k] = \vert {\cal
A}\vert $. \medskip

By applying Theorem \ref{fMirsky} to the poset $P_G$, we get the
following.

\Corollary \label{fcovering} The minimum number of stable matchings of
a bipartite preference system covering a non-negative integer-valued
function $f$ on $E_{\rm st}$ is equal to the maximum total $f$-value
of an anti-stable set of edges.  In particular, the minimum number of
stable matchings covering all stable edges is equal to the maximum
cardinality of an anti-stable set.  Furthermore, with the help of the
two-phase greedy algorithm concerning the weighted Mirsky problem,
both a minimum family of stable matchings covering $f$ (which contains
at most $\vert E_{\rm st}\vert $ distinct members) and an anti-stable
set with maximum total $f$-value can be computed in strongly
polynomial time.  \FBOX \eCo

\subsubsection{Weighted Dilworth} \label{WD}

Let $P=(S,\preceq )$ be a poset.  It is well-known (and follows from
Dilworth's theorem) that the graph defined by $P$ (in which $uv$ is an
edge if $u$ and $v$ are comparable) is perfect.  The first part of the
next proposition follows from Theorem 5 of the classic paper of Lovász
\cite{Lovasz72}, while the second part is a consequence of Theorem 4 of
the paper \cite{CFS} of Cook, Fonlupt, and Schrijver (stating that in a perfect graph $G=(V,E)$ endowed with a non-negative integer-valued weight-function $w$ on $V$, there exists a smallest system of cliques covering $w$ in which the number of distinct cliques is at most $|V|$).

\Proposition \label{CFS} The polytope of antichains of a poset
$P=(S,\preceq )$ is described by the following linear system.  $$
\hbox{ $\{x\in {\bf R}_+\sp S:  \ \widetilde x(C)\leq 1$ \ for every
(maximal) chain $C$ of $P\}$.}\ $$ Moreover, this system is TDI, and the dual has the integer Caratheodory property in the following sense: for any non-negative integer-valued weight function $w$, an optimal integer dual solution exists whose support consists of linearly independent chains. \FBOX \eP

\medskip

We say that a family $\cal C$ of chains {\bf covers} a non-negative
weight-function $w:  S\rightarrow {\bf Z} _+$ if every element $s\in
S$ occurs in at least $w(s)$ members of $\cal C$.  The following
theorem is a direct consequence of Proposition \ref{CFS}, where Part
{\bf (A)} is nothing but the weighted extension of Dilworth's theorem
(see, for example, Theorem 14.3 in \cite{Schrijverbook}).

\THEOREM \label{wDilworth} \noindent {\bf (A)} {\bf (weighted
Dilworth)} \ \ Given a non-negative integer-valued function $w$ on the
ground-set of a poset $P=(S,\preceq )$, the maximum $w$-weight of an
antichain of $P$ is equal to the minimum number of chains covering
$w$.  If $P$ is D-antichain-extendible, then the maximum $w$-weight
antichain may be chosen to be a D-antichain.

\noindent {\bf (B)} \ The minimizer $w$-covering family of chains can
be chosen in such a way that the number of distinct chains is at most
$\vert S\vert $. \FBOX \eT

Part {\bf (A)} follows easily from the original theorem of Dilworth if
we replace each element $s\in S$ by $w(s)$ elements which form an
antichain and the new elements have the same relationship to other
elements as $s$ has.  This approach, however, does not provide a
polynomial time algorithm for computing a maximum $w$-weight antichain
(the primal problem) and a minimum cover of $w$ by chains (the dual
problem).

By applying Proposition \ref{CFS} and Theorem \ref{wDilworth} to poset
$P_G$, we get the following.

\Corollary \label{wcovering} Consider a preference system on a
bipartite graph $G=(U,W;E)$ in which each stable matching is a perfect
matching.  A linear system describing the polytope of stable matchings
is as follows.  $$ \hbox{ $\{x\in {\bf R}_+\sp {E_{\rm st}} :  \
\widetilde x(E_{\rm st}) =\vert U\vert $ \ and $ \widetilde x(A)\leq
1$ \ for every (maximal) anti-stable set $A\subseteq E_{\rm st}\}.$ }\
$$ This system is TDI.  For a non-negative integer-valued
weight-function $w$ on $E_{\rm st}$, the maximum $w$-weight of a
stable matching of a preference system defined on $G$ is equal to the
minimum number of anti-stable sets covering $w$.  The smallest
$w$-covering family of anti-stable sets can be chosen in such a way
that it contains at most $\vert E_{\rm st}\vert $ distinct anti-stable
sets.  In particular, a given set $H$ of stable edges includes a
stable matching if and only if $H$ cannot be covered by less than
$\vert W\vert $ anti-stable sets.  \FBOX \eCo

\Remark \label{TDI2} {\em As we already indicated in Remark
\ref{TDI1}, Rothblum \cite{Rothblum92} provided a linear description
of the polytope of stable matchings, which uses only $O(\vert E\vert
)$ linear inequalities, and this system was shown to be TDI by Kir\'aly
and Pap \cite{Kiraly+Pap}.  It is an interesting challenge to derive
the TDI-ness of the Rothblum system from Corollary \ref{wcovering}.
$\bullet $ }\eRe

\subsubsection{Weighted Dilworth algorithmically}

Our next goal is to provide a constructive proof of
Theorem \ref{wDilworth} by describing a strongly polynomial algorithm
that solves both the primal and the dual problem. First, we give an algorithmic proof of Part {\bf (A)} of Theorem \ref{wDilworth}.  The approach may be viewed as a weighted extension of Fulkerson's
\cite{Fulkerson56} elegant proof for the Dilworth theorem, which is
based on a reduction to K{\H o}nig's theorem. This algorithm
proves a weaker form of Part {\bf (B)}  by providing an
optimal family of chains that consists of at most $4|S|$ distinct chains. We will show at the end of the subsection that we can obtain a family of at most $|S|$ distinct chains algorithmically, and thus prove Part {\bf (B)}, by calling the above algorithm as a subroutine at most $2|S|$ times.

Let $b:V\rightarrow {\bf Z}_+$ be a non-negative integer-valued
function on the node-set of a bipartite graph.  A function
$z:E\rightarrow {\bf Z}_+$ is called a {\bf $b$-matching} if $d_z(v)
\leq b(v)$ holds for every node $v$ of $G$.  (Note that $z$ may have
components larger than $1$.)  We need the following extension of K{\H
o}nig's theorem (see, for example, Theorem 21.1 in the book
\cite{Schrijverbook} of Schrijver in its special case $w\equiv 1$.)

\Lemma \label{bmatching} Let $G=(V,E)$ be a bipartite graph and
$b:V\rightarrow {\bf Z_+} $ a function on its node-set.  Then $$
\hbox{ $\max \{\widetilde z(E):  \ z$ \ a $b$-matching of \ $G \} =$
}\ $$ $$ \hbox{ $\min \{\widetilde b(L) :  \ L\subseteq V$ \ a
covering of \ $E\}$.  }\ $$ The optimal $b$-matching can be chosen in
such a way that the set of $z$-positive edges forms a forest, and
hence it consists of at most $\vert V\vert -1$ edges.  \eL

\Proof The min-max formula follows immediately from the linear
programming duality theorem and from the fact that the
node-edge-incidence matrix of a bipartite graph is totally unimodular.

To see the second part of the lemma, let $z$ be an optimal
$b$-matching and suppose that there is a circuit $C=\{e_1,e_k,\dots
,e_{2k}\}$ consisting of $z$-positive edges.  Let $\alpha $ denote the
minimum of these $z$-values, and suppose that this minimum is attained
on edge $e_1$.

Decrease the $z$-values on the edges of $C$ with odd subscript by
$\alpha $ and increase the $z$-values on edges with even subscript by
$\alpha $. Let $z'$ denote the modified vector.  Then $z'$ is also an
optimal $b$-matching for which the number of $z'$-positive edges is
smaller than the number of $z$-positive edges.  \FBOX \medskip

It is also a well-known fact that both a largest $b$-matching and a
minimum $b$-weight covering of the edge-set can be computed in
strongly polynomial time by applying a standard MFMC algorithm.

\medskip

\noindent \textbf{Algorithmic proof of Part (A) of Theorem \ref{wDilworth}.} Let us turn to the algorithmic proof of the non-trivial direction
$\max\geq \min$ of the min-max formula in Theorem \ref{wDilworth}.
We may assume that $w$ is strictly positive since if $w(s)=0$ for some
element $s\in S$, then the removal of $s$ from $P$ affects neither the
maximum weight of antichains nor a family of chains covering $w$.

We assign a bipartite graph $G_P=(S',S'';E_P)$ to poset $P=(S, \preceq
)$, where $S'$ and $S''$ are disjoint copies of $S$, and $u'v''$ is an
edge if $u\succ v$.  Apply Lemma \ref{bmatching} to the function $b$
defined on the node-set of $G_P$ where $$b(s'):=b(s''):= w(s) \ \ \ \
(s\in S).$$

Let $L_0$ \ ($\subseteq S'\cup S''$) be a minimum $b$-value covering
of the edge-set of $G_P$, and let $z_0$ an optimal $b$-matching of
$G_P$, for which the set of $z_0$-positive edges form a forest.  Then
the number of $z_0$-positive edges is at most $2|S|$.
By Lemma
\ref{bmatching}, we have $$\widetilde b(L_0) = \widetilde z_0(E_P).$$

Since $w$ (and hence $b$, as well) is strictly positive, the set $L_0$
is an inclusionwise minimal covering of $E_P$.  It is not possible for
an element $s\in S$ that both $s'$ and $s''$ belong to $L_0$.  Indeed,
if $s'\in L_0$, then it follows from the minimality of $L_0$ that there
is an element $t_1\in S$ for which $s\succ t_1$ and $t_1''\not \in
L_0$, and analogously, if $s''\in L_0$, then there is an element
$t_2\in S$ for which $t_2 \succ s$ and $t_2''\not \in L_0$.  But then
$t_2\succ t_1$, and hence $t_2't_1''$ is also an edge of $G_P$, which
is not covered by $L_0$.  Therefore, it is indeed not possible for
both $s'$ and $s''$ to be in $L_0$.

Let $A_* := \{s\in S:  \ s', s'' \not \in L_0\}$.  Then $A_*$ is an
antichain of $P$ for which

\eq \widetilde w(A_*) = \widetilde w(S) - \widetilde b(L_0) =
\widetilde w(S) - \widetilde z_0(E_P).  \label{(wA0)} \eeq

\medskip

Consider the acyclic digraph $D_P=(V_P, A_P)$ where $V_P :=S\cup
\{s,t\}$ and $$ \hbox{$A_P := \{uv:  u,v \in S , u\succ v\} \ \cup \
\{sv:  v\in S \} \ \cup \ \{vt:  v\in S \}.$ }\ $$

Define the function $z_* :  A_P\rightarrow {\bf Z}_+$ as follows,

$$ \hbox{ $z_*(uv):  = z_0(u'v'')$ \ if \ $u,v\in S$ and $u\succ v$,
}\ $$ $$ \hbox{ $z_*(sv):  = w(v) -d_{z_0} (v'')$ \ if \ $v\in S$, }\
$$ $$ \hbox{ $z_*(vt):  = w(v) -d_{z_0} (v')$ \ if \ $v\in S$.  }\ $$

It can be easily seen that $z_*$ forms an $st$-flow which is positive
on at most $4|S|$
arcs.  By \eref{(wA0)}, the flow-amount of $z_*$ is
\eq \delta _{z_*}(s)= \sum _{v\in S} [ w(v) -d_{z_0}(v') ] =
\widetilde w(S) - \widetilde z_0(E_P) = \widetilde w(A_*).
\label{(wAz)} \eeq

It is a well-known property of flows that an arbitrary non-negative
integer-valued flow $z$ which is positive on $\ell$ arcs can be
produced in a greedy way as the sum of at most $\ell$ path-flows
(where a path-flow is a constant integer along an $st$-path and $0$
otherwise).

Therefore, $z_*$ can be obtained as the sum of at most $4|S|$
path-flows.
Since $\varrho _{z_*}(v) = w(v) = \delta _{z_*}(v)$ holds for every
node $v\in S$ of the digraph $D_P$, by restricting these path-flows to
$S$, we obtain a family of chains of $P$ covering $w$ and consisting
of $\delta _{z_*}(s)$ chains, in which the number of distinct chains
is at most $4|S|$.  \FBOX

\medskip

Before we give an algorithm for Part \textbf{(B)}, we prove a claim about a cone defined by the chains in a poset, which is an easy consequence of Proposition \ref{CFS}.

\Claim \label{cl:cone}
Let $P=(S,\preceq )$ be a poset, and let $K \subseteq \mathbf{R}^{|S|+1}$ denote the cone generated by the vectors $\{(1,\chi^C): \text{ $C$ is a chain of $P$}\}$. Then
\[
    K=\{(t,x): t \in \mathbb{R}_+,\ x \in \mathbb{R}^{S}_+,\ \widetilde{x}(A) \leq t \text{ for every antichain $A$ of $P$}\}.
\]
\eCl

\Proof Since $|A \cap C| \leq 1$ for every chain $C$ and antichain $A$, any $(t,x)\in K$ satisfies $\widetilde{x}(A)\leq t$ for every antichain $A$. Conversely, if $t>0$ and $(t,x)$ satisfies $\widetilde{x}(A)\leq t$ for every antichain $A$, then $\widetilde{x}(A)/t \leq 1$ for every antichain $A$, so $x/t$ is a convex combination of characteristic vectors of cliques by Proposition \ref{CFS}, and therefore $(t,x)\in K$. The case $t=0$ is obvious because only the all-zero vector $x\equiv 0$ satisfies $\widetilde{x}(A)\leq 0$ for every antichain $A$. 
\FBOX

\medskip

\noindent \textbf{Algorithmic proof of Part (B) of Theorem \ref{wDilworth}.}
To obtain an optimal family of chains of $P$ covering $w$ in which the number of distinct chains is at most $|S|$ as required by Part \textbf{(B)} of Theorem \ref{wDilworth}, we can rely on a greedy method that was implicitly described in \cite{CFS} and \cite{Sebo90}. Let $\gamma$ be the maximum $w$-weight of an antichain of $P$, let $A_0$ be a maximum weight antichain, and let $\mathcal{C}$ be a family of chains covering $w$ such that $|\mathcal{C}|=\gamma$. These can be found using the algorithm for Part (\textbf{A}) (it may also be assumed that the number of distinct chains in $\mathcal{C}$ is at most $4|S|$, but we will not use this fact). We can observe that $(\gamma,w)\in K$, where $K$ is the cone defined in Claim \ref{cl:cone}.

Let $C_1$ be an arbitrary member of $\mathcal{C}$. If $\widetilde{w}(A)=\gamma$ for some antichain $A$, then $|A \cap C_1|=1$ by complementary slackness.
 
Let $\gamma_1$ be the maximum $w$-weight of an antichain disjoint from $C_1$, and let $A_1$ be an antichain disjoint from $C_1$ with $\widetilde{w}(A_1)=\gamma_1$; these can be computed using the algorithm for Part (\textbf{A}). Let $\lambda_1:=\gamma-\gamma_1$; note that $\lambda_1$ is positive by complementary slackness, and it is integer since both $\gamma$ and $\gamma_1$ are integers.  

We consider the weight function $w_1 :=(w-\lambda_1 \chi^{C_1})^+$. Observe that $\widetilde{w_1}(A) \leq \gamma_1$ for every antichain $A$, because $\widetilde{w}(A)\leq \gamma$ and $|C_1 \cap A|\leq 1$, so $\widetilde{w_1}(A) \leq \gamma-\lambda_1=\gamma_1$. Furthermore, 
we claim that if $\widetilde{w}(A)=\gamma$, then $\widetilde{w_1}(A)=\gamma_1$. Indeed, in this case $|C_1 \cap A|= 1$ by complementary slackness, so $\widetilde{w_1}(A) = \gamma-\lambda_1=\gamma_1$.

The above observations imply that $(\gamma_1,w_1) \in K$, and $(\gamma_1,w_1)$ is on the smallest face of $K$ containing $(\gamma,w)$. Furthermore, $\widetilde{w_1}(A_1)=\gamma_1$ but $\widetilde{w}(A_1)<\gamma$, so $(\gamma_1,w_1)$ is on a proper subface of the smallest face containing $(\gamma,w)$. For the purposes of the following argument, we will use the notation $\gamma_0=\gamma$ and $w_0=w$.

Suppose that we have already computed $C_i$, $\gamma_i$ and $w_i$. If $w_i \equiv 0$, we stop; otherwise, let $C_{i+1}$ be a chain in an optimal family of chains covering $w_i$, which can be computed using the algorithm for Part (\textbf{A}). Let $\gamma_{i+1}$ be the maximum $w_i$-weight of an antichain disjoint from $C_{i+1}$, let $\lambda_{i+1}:=\gamma_i-\gamma_{i+1}$, and let $w_{i+1}=(w_i-\lambda_{i+1} \chi^{C_{i+1}})^+$. By the same argument as in the $i=0$ case, $\lambda_{i+1}$ is a positive integer, and $(\gamma_{i+1},w_{i+1})$ is on a proper subface of smallest face of $K$ containing $(\gamma_i,w_i)$.

The above procedure produces chains $C_1,\dots, C_k$ and positive integer coefficients $\lambda_1,\dots,\lambda_k$ such that $\sum_{i=1}^k \lambda_i=\gamma$ and $\sum_{i=1}^k \lambda_i\chi^{C_i}\geq w$. Furthermore, $k\leq |S|$ since the dimension of the smallest face of $K$ that contains $(\gamma_i,w_i)$ strictly decreases in each step, and $(\gamma_0,w_0)$ is on a face of dimension at most $|S|$ because $\widetilde{w_0}(A_0)=\gamma_0$. 

The property that the dimension of the smallest face containing $(\gamma_i,w_i)$ decreases in each step also implies that the vectors $(1,\chi^{C_1}), \dots, (1,\chi^{C_k})$ are linearly independent, because $(\gamma_{i-1},w_{i-1})-(\gamma_{i},w_{i})=\lambda_{i} \cdot(1, \chi^{C_i})$ for every $i$. Thus, the vectors $\chi^{C_1}, \dots, \chi^{C_k}$ are affinely independent. To show that they are in fact linearly independent, we can observe that $|C_i \cap A_0|=1$ for every $i\in \{1\dots,k\}$ by complementary slackness, so any linear dependence between the vectors $\chi^{C_1}, \dots, \chi^{C_k}$ must be an affine dependence.
\BB

\subsection{The theorem of Greene+Kleitman}

A fundamental theorem of Greene+Kleitman \cite{Greene-Kleitman}
provided a profound extension of Dilworth' theorem by formulating and
proving an elegant min-max formula for the maximum cardinality $\alpha
_\ell$ of the union of $\ell$ antichains.  For a convenient and
concise optimality criterion, the concept of orthogonality was
introduced in \cite{FrankJ5}.  A family $\cal A$ of disjoint
antichains and a family $\cal C$ of disjoint chains are called {\bf
orthogonal} if $S= (\cup {\cal A}) \ \cup \ (\cup {\cal C})$ and each
member of ${\cal A}$ intersects (in one element) each member of $\cal
C$.

\THEOREM [Greene+Kleitman] \label{GK} In a poset $P=(S,\preceq )$, the
maximum cardinality $\alpha _\ell$ of the union of $\ell$ antichains
is equal to the minimum of $$ \ell \vert {\cal C}\vert + \vert S -
\cup {\cal C}\vert $$ \noindent where the minimum is taken over all
families $\cal C$ of disjoint chains.  A family ${\cal A}_\ell$ of
$\ell$ disjoint antichains has a maximum cardinality union if and only
if there exists a family of disjoint chains orthogonal to ${\cal
A}_\ell$.  \FBOX \eT

In the paper of Frank \cite{FrankJ5} (see also Section 3.6.2 of book
\cite{Frank-book}), a strongly polynomial algorithm, based on the
min-cost flow algorithm of Ford and Fulkerson, was described to
compute a family ${\cal A}_\ell$ of $\ell$ antichains whose union has
a maximum number of elements, along with a family of chains which is
orthogonal to ${\cal A}_\ell$.  The theorem of Greene+Kleitman
immediately implies the following.

\Corollary \label{diszdanti} In a poset $P=(S,\preceq )$, the maximum
number of disjoint $D$-antichains is equal to the minimum number of
elements intersecting all $D$-antichains.  \FBOX \eCo

The algorithm in \cite{FrankJ5}, when specialized to this case,
computes in strongly polynomial time both the largest set of disjoint
$D$-antichains and the smallest set of elements intersecting all
$D$-antichains.

In a D-antichain-extendible poset, the maximum cardinality of the
union of $\ell$ antichains is the same as the maximum cardinality of
the union of $\ell$ D-antichains, and therefore Theorem \ref{GK}
provides a min-max theorem for this case.  Note, however, that the
$\ell$ antichains in Theorem \ref{GK} with a largest union can
trivially be chosen to be pairwise disjoint, while in the case of the
D-antichain packing problem, this cannot be an expectation.
Therefore, it is useful to extend the concept of orthogonality to the
case when the family of antichains may have non-disjoint members, as
follows.  A family $\cal A$ of (not necessarily disjoint) antichains
and a family $\cal C$ of disjoint chains are {\bf orthogonal} if

$$ \begin{cases} & S= (\cup {\cal A}) \ \cup \ (\cup {\cal C}),

\\ & \hbox{each member of ${\cal A}$ intersects (in one element) each
member of $\cal C$, }\

\\ & \hbox{the members of $\cal A$, when restricted to $\cup {\cal
C}$, are disjoint.  }\

\end{cases} $$

\medskip

With this notion, Theorem \ref{GK} transforms into the following.

\THEOREM \label{GKDa} In a D-antichain-extendible poset $P=(S,\preceq
)$, the maximum cardinality $\alpha _\ell$ of the union of $\ell$
D-antichains is equal to the minimum of

$$ \ell \vert {\cal C}\vert + \vert S - \cup {\cal C}\vert $$

\noindent where the minimum is taken over all families $\cal C$ of
disjoint chains.  A family ${\cal A}_\ell$ of $\ell$ disjoint
D-antichains has a maximum cardinality union (that is, $\vert \cup
{\cal A}_\ell\vert = \alpha _\ell$) if and only if there exists a
system of disjoint chains orthogonal to ${\cal A}_\ell$.  \FBOX \eT

Let $G=(U,W;E)$ be a bipartite graph endowed with a preference system.
Let ${\cal M}_\ell$ be a family of $\ell$ not necessarily disjoint
stable matchings, and let ${\cal K}$ be a system of disjoint
anti-stable sets.  We say that ${\cal M}_\ell$ and ${\cal K}$ are {\bf
orthogonal} if

$$ \begin{cases} & E_{\rm st} = \cup {\cal M}_\ell \ \cup \ \cup {\cal
K},

\\ & \hbox{each member of ${\cal M}_\ell$ intersects (in one element)
each member of $\cal K$, }\

\\ & \hbox{the members of ${\cal M}_\ell$, when restricted to $\cup
{\cal K}$, are disjoint.  }\

\end{cases} $$

By applying Theorem \ref{GKDa} to the $G$-induced poset $P_G$, we
obtain the following.

\Corollary \label{anti-stable} Given a preference system on a
bipartite graph $G$, the maximum cardinality of the union of $\ell$
(non-necessarily disjoint) stable matchings is equal to the minimum of

\eq \ell \vert {\cal K} \vert + \vert E_{\rm st} - \cup {\cal K} \vert
\eeq

\noindent where $\cal K$ is a system of disjoint anti-stable sets.  A
family ${\cal M}_\ell$ of $\ell$ stable matchings has a union with
maximum cardinality if and only if there is a system of disjoint
anti-stable sets which is orthogonal to ${\cal M}_\ell$.  \FBOX \eCo

\medskip In Section \ref{pack2.2}, we described an algorithm for the
weighted extension of this last problem when, given a weight-function
$w$, we wanted to find $\ell$ (non-necessarily disjoint) stable
matchings for which the $w$-weight of their union is maximum.

\subsection{An algorithm for packing D-antichains}

Corollary \ref{fcovering} provided a min-max formula for the minimum
number of stable matchings covering a lower-bound function $f$.  Its proof
relied on a two-phase greedy algorithm concerning the weighted
Mirsky's theorem.  In this section, we investigate the packing
counter-part of this problem, when for a given upper-bound function $h$,
we want to find a maximum number of stable matchings such that each
edge $e$ belongs to at most $h(e)$ of them.

Let $P=(S,\preceq )$ be again a poset and let $h:S\rightarrow {\bf Z}
_+$ be a non-negative integer-valued upper-bound function.  Let $\alpha
:=\alpha _P$ denote the cardinality of a $D$-antichain, while ${\cal
D} :={\cal D}_P$ is the set of D-antichains of $P$.  We say that a
family of D-antichains is {\bf $h$-independent} if every element $s\in
S$ belongs to at most $h(s)$ members of the family.  A subset
$B\subseteq S$ {\bf blocks} (or is a {\bf blocker} of) $\cal D$ if it
intersects all members of $\mathcal D$.  For example, (by Dilworth's
theorem) any chain in a smallest chain-decomposition of $P$ is a
blocker of $\cal D$.  It should, however, be noted that there exists a
D-antichain-extendible poset in which no smallest blocker of
D-antichains is a chain.

\THEOREM \label{Danti} Given a non-negative integer-valued function
$h$ on the ground-set $S$ of a poset $P$, the maximum cardinality of
an $h$-independent family of D-antichains is equal to the minimum
$h$-value of a blocker $B\subseteq S$ of D-antichains.  A largest
$h$-independent family of antichains can be chosen in such a way that
the number of distinct antichains is at most $\vert S\vert $. \eT

We remark that the theorem for $h\equiv 1$ is a special case (or
consequence) of Theorem \ref{GK} of Greene and Kleitman.  For general $h$,
the min-max formula follows from this if we replace each element $s\in
S$ by a chain of $h(s)$ new elements.  Since this approach is not
polynomial in $h$, we show how the algorithm described in the proof of
Theorem \ref{pakol} can be used.

\medskip

\noindent {\bf Algorithmic proof of Theorem \ref{Danti}.} \ \ Let
$\{C_1,C_2,\dots ,C_\alpha \}$ be a Dilworth decomposition of $P$ into
chains where $\alpha $ denotes the cardinality of a largest
antichain of $P$.  Define a digraph $D=(V,A)$ in which $s\sp *,t\sp
*\in V$ and $D$ consists of $\alpha $ openly disjoint $s\sp *t\sp
*$-paths $P_1,\dots ,P_\alpha $. Here $P_i$ has $\vert C_i\vert $ arcs,
and the arcs of $P_i$ correspond to the elements of $C_i$ in such a
way that an arc $e$ of $P_i$ precedes another arc $f$ of $P_i$ if the
element of $C_i$ corresponding to $e$ is larger than the element of
$C_i$ corresponding $f$.  Let $h(e)$ be the $h$-value of the element
of the poset corresponding to $e$.

For a D-antichain $A'$, we associate with $A'$ the set of nodes of $D$
which precede the $\alpha $ arcs of $D$ corresponding to the elements
of $A'$.  (In particular, this means the singleton $\{s\sp *\}$ is
associated with the unique highest D-antichain of $P$, while the set
$V-t\sp *$ is associated with the lowest D-antichain.)

The system of subsets of $V$ associated with the D-antichains of $P$
form a ring-set ${\cal R}_0$.  By applying Theorem \ref{pakol} and its
algorithmic proof to this special digraph and ring-set, we obtain
Theorem \ref{Danti} as well as an algorithm to compute a largest
family $\cal F$ of $h$-independent D-antichains of poset $P$ along
with a blocker $B$ of D-antichains for which $\widetilde h(B)$ is
minimum (that is, $\vert {\cal F}\vert = \widetilde h(B)$).  \FBOX

\Remark {\em Suppose that in Theorem \ref{Danti} we are also given a
non-negative integer-valued weight-function $w$ on $S$.  It is an easy
exercise to prove that not only the D-antichains are closed under the
meet and join operations, but the maximum $w$-weight D-antichains as
well.  Therefore the algorithmic proof of Theorem \ref{Danti} outlined
above can be easily extended to one that computes a largest family of
$h$-independent maximum $w$-weight antichains along with a blocker $B$
of maximum $w$-weight D-antichains for which $\widetilde h(B)$ is
minimum.

When we specialize this to the $G$-induced poset $P_G$, we obtain a
min-max formula for the maximum number of $h$-independent maximum
$w$-weight stable matchings, and this result is just equivalent to
Theorem \ref{mincostpak2}.  $\bullet $ }\eRe

In the special case $h\equiv 1$ of Theorem \ref{Danti}, we are back at
Corollary \ref{gindep}.  We emphasize, however, that Theorem
\ref{Danti} can be applied to matroid kernels as well \cite{FFK2}.

\subsubsection{Weighted Greene+Kleitman}

Let $w:S\rightarrow {\bf Z} _+$ be a weight-function.  For a
$w$-independent family $\cal C$ of chains, we say that an element
$s\in S$ is {\bf unsaturated} if it is contained in less than $w(s)$
members of $\cal C$.  For $s\in S$, let

$$ \hbox{ $\sigma _{\cal C}(s) := w(s) - \vert \{C\in \cal C:  \ s\in
C\}\vert $ }\ .$$ By a standard element-multiplication technique, the
Greene+Kleitman theorem immediately implies its weighted extension.

\THEOREM \label{wGK} Let $w$ be a non-negative integer-valued
weight-function on the ground-set of poset $P=(S,\preceq )$.  The
maximum $w$-weight $\alpha _\ell(w)$ of the union of $\ell$ antichains
is equal to the minimum of $$ \ell \vert {\cal C}\vert + \sum _{s\in
S} \sigma _{\cal C}(s) ,$$

\noindent where the minimum is taken over all $w$-independent families
$\cal C$ of chains.  There is a minimizer $w$-independent family $\cal C$ of chains in which the number of its distinct members is at most $\vert S\vert$. A family ${\cal A}_\ell$ of $\ell$ disjoint antichains has a
maximum $w$-weight union if and only if there exists a $w$-independent
family $\cal C$ of chains which is orthogonal to ${\cal A}_\ell$ in
the sense that \ {\bf (A)} \ $\cup {\cal A}_{\ell}$ contains each
element unsaturated by $\cal C$ and \ {\bf (B)} \ each member of
${\cal A}$ intersects (in one element) each member of $\cal C$.  \FBOX
\eT

\subsubsection{Disjoint D-antichains with cheapest union}

For an application in a subsequent work \cite{FFK2}, we show how the
approach in Corollary \ref{main1b} can be applied to constructing
$\ell$ disjoint D-antichains of a poset $P=(S,\preceq )$ endowed with
a cost-function $c:S\rightarrow {\bf Z} _+$ for which the $c$-cost of
their union is minimum.

Let $\alpha $ denote the cardinality of a D-antichain.  By Dilworth,
there is a partition $\{C_1,\dots ,C_\alpha \}$ of $S$ into $\alpha $
chains.  Let $D$ be a digraph consisting of $\alpha $ openly disjoint
$st$-paths, where the arcs of path $P_i$ correspond to the elements of
$C_i$.  For an arc $a$ of $D$ let $g(a)$ be the $c$-cost of the
corresponding element of $S$.

For a D-antichain $A$ of $P$, let $A_-:= \{u:  u\preceq v$ for some
$v\in A\}$ denote the lower ideal of $A$, and let $A'$ denote the arcs
of $D$ corresponding to the elements of $A$.  Now the set of nodes of
$D$ preceding $A'$ corresponds to $A_-$, and these sets associated
with the D-antichains of $P$ form a ring-family $\cal R$.  It follows
that Corollary \ref{main1b}, when applied to this case, results in an
$\cal R$-compatible $\ell$-cut $L$ of $D$ whose $g$-value is minimum,
and this $L$ defines a family of $\ell$ disjoint D-antichains of $P$
for which the $c$-cost of their union is minimum.

\medskip This algorithm for computing $\ell$ disjoint D-antichains can
easily be used for

\ $(*)$ \ finding $\ell$ not necessarily disjoint D-antichains whose
union is of maximum $w$-weight.

To this end, replace each element $s\in S$ by a chain of $\ell $
elements, where the weight of the first element of the chain is $w(s)$
and zero of the others.  Then $\ell$ disjoint D-antichains with
maximum total weight determines an optimal solution to $(*)$.

{\bf Acknowledgement} \ \ We are grateful to \'Agnes Cseh, P\'eter
Madarasi, David Manlove, and Andr\'as Seb{\H o} for their invaluable
comments and suggestions.

\end{document}